\documentclass[11pt,letterpaper]{article}
\usepackage{fullpage}
\usepackage{amsmath,amssymb,amsthm,amsfonts,graphicx}

\newcommand{\sve}{\vec{s}\,}
\newcommand{\svp}{\vec{s}\,'}
\newcommand{\Mlim}{M\rightarrow \infty}
\newcommand{\ep}{\epsilon}
\newcommand{\norms}[1]{\| #1\|}        
\newcommand{\ket}[1]{\left| #1\right\rangle}        
\newcommand{\bra}[1]{\left\langle #1\right|}        
\newcommand{\kets}[1]{| #1\rangle}        
\newcommand{\bras}[1]{\langle #1|}        
\newcommand{\braket}[2]{\langle #1 | #2 \rangle} 

\newcommand{\tr}{{\textrm{tr}\,}}
\newcommand{\means}[1]{\langle #1 \rangle}        
\newcommand{\corrf}{\means{ \sigma^{(i)}_z \sigma^{(j)}_z } 
- \means{ \sigma^{(i)}_z }\means{ \sigma^{(j)}_z }}
\newcommand{\corrfO}{\means{ O^{(i)} O^{(j)} } - \means{ O^{(i)} }\means{ O^{(j)} }}

\begin{document}


\title{The Quantum Transverse Field Ising Model \\ 
on an Infinite Tree from Matrix Product States.}
\author{Daniel Nagaj\thanks{nagaj@mit.edu}, Edward Farhi, Jeffrey Goldstone, Peter Shor and Igor Sylvester\\
MIT Center for Theoretical Physics}

\date{\today}

\maketitle

\begin{abstract}
We give a generalization to an infinite tree geometry of Vidal's  
infinite time-evolving block decimation (iTEBD) algorithm \cite{Vidal1Dinfinite}
for simulating an infinite line of quantum spins. 
We numerically investigate the quantum Ising model in a transverse field on the Bethe lattice using the Matrix Product State ansatz. We observe a second order phase transition,
with certain key differences from the transverse field Ising model on an infinite spin chain.
We also investigate a transverse field Ising model with a specific longitudinal field. When the transverse field is turned off, this model has a highly degenerate ground state
as opposed to the pure Ising model whose ground state is only doubly degenerate.
\end{abstract}


\section{Introduction}

The matrix product state (MPS) description \cite{OstlundRommerMPS} has brought a new way of approaching many-body quantum systems. Several methods of investigating spin systems have been developed recently combining state of the art many-body techniques 
such as White's Density Matrix Renormalization Group \cite{DMRG}\cite{DMRGreview} (DMRG) with quantum information motivated insights. Vidal's Time Evolving Block Decimation (TEBD) algorithm 
\cite{VidalMPS} \cite{VidalMPS2} uses MPS and emphasizes entanglement (as measured by the Schmidt number), directing the computational resources into that bottleneck of the simulation. It provides the ability to simulate time evolution and it was shown that MPS-inspired methods handle periodic boundary conditions well in one dimension \cite{DMRGandPBC}, areas where the previous use of DMRG was limited.
TEBD has been recast into the language of DMRG in \cite{TEBDandDMRG}
and adapted to finite systems with tree geometry in \cite{ShiTrees}. 
DMRG is especially successful in describing the properties of quantum spin chains, the application of basic DMRG-like methods is limited for quantum systems with higher dimensional geometry. New methods like PEPS \cite{PEPS} generalize MPS to higher dimensions, opening ways to numerically investigate systems that were previously inaccessible.

We are interested in investigating infinite translationally invariant systems.
Several numerical methods to investigate these were developed recently.
The iTEBD algorithm \cite{Vidal1Dinfinite} (see also Sec.\ref{TIsection}) is a generalization of TEBD to infinite one-dimensional systems. A combination of PEPS with iTEBD called iPEPS \cite{iPEPS} provides a possibility of investigating infinite translationally invariant systems in higher dimensions.
\begin{figure}
	\begin{center}
	\includegraphics[width=1.1in]{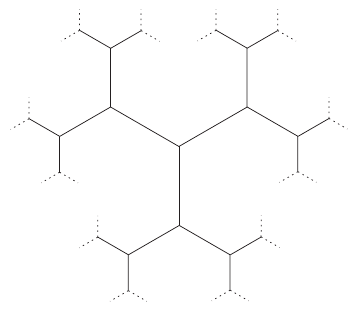}
	\caption{The Bethe lattice (infinite Cayley tree).}
	\label{figinfitree}
	\end{center}
\end{figure}
Our contribution is a method to investigate the ground state properties of infinite translationally invariant quantum systems on the Bethe lattice using imaginary time evolution with Matrix Product States. The Bethe lattice is an infinite tree with each node having three neighbors, as depicted in Fig.\ref{figinfitree}. It is translationally invariant in that it looks the same at every vertex.
This geometry is interesting, because of the following connection to large random graphs with fixed valence. Moving out from any vertex in such a random graph, 
you need to go a distance of order $\textrm{log}\,n$, where $n$ is the number of vertices in the graph, before you detect that you are not on the Bethe lattice, that is, before you see a loop.

We choose to investigate the quantum transverse field Ising model on the Bethe lattice. 
Note that we work directly on the infinite system, never taking a limit.
First we test the iTEBD method on a system with a known exact solution, the infinite line. Then we turn to the Bethe lattice with the new method we provide. In both cases, the Hamiltonian is given by
\begin{eqnarray} 
	H = \frac{J}{2} \sum_{\langle i,j\rangle} (1-\sigma_z^{i} \sigma_z^{j}) 
		+ \frac{h}{2} \sum_i \left(1-\sigma_x^{i}\right), \label{ourH1}
\end{eqnarray}
where the sum over $i$ is over all sites, and the sum over $\langle i,j \rangle$ is over all bonds (nearest neighbors).
We show that imaginary time evolution within the MPS ansatz provides a very good approximation for the exact ground state on an infinite line, giving us nearly correct critical exponents for the magnetization and correlation length as we approach the phase transition.
We obtain new results for the quantum Ising model in transverse field on the infinite tree.
Similarly to the infinite line, we observe a second order phase transition and 
obtain the critical exponent for the magnetization, $\beta_{T}\approx 0.41$ (different than the mean-field result). However, the correlation length does not diverge at the phase transition for this system and we conjecture that it has the value $1/\ln 2$. 

We also investigate a model where besides an antiferromagnetic interaction of spins we add a specific longitudinal field $\frac{1}{4}\sigma_z^i$ for each spin: 
\begin{eqnarray}
	H_{\textsc{not}\,00} 
	&=& J 
	 \sum_{\langle i,j\rangle} 
		\frac{1}{4}\left( 1 + \sigma_z^i + \sigma_z^j + \sigma_z^i \sigma_z^j  \right)
		+ \frac{h}{2} \sum_i \left(1-\sigma_x^{i}\right).
		\label{ourH00}
\end{eqnarray}
We choose the longitudinal field in such a way that the interaction term in the computational basis takes a simple form, $\ket{00}\bra{00}_{ij}$, giving an energy penalty to the $\ket{00}$ state of neighboring spins. (We follow the usual convention that spin up in the $z$-direction is called 0.) We call it the \textsc{not} 00 model accordingly. This model is interesting from a computational viewpoint. 
The degeneracy of the ground state of $H_{\textsc{not}\,00}$ at $h=0$ is high for both infinite line and infinite tree geometry of interactions. We are interested in how our numerical method deals with this case, as opposed to the double degeneracy of the ground state of \eqref{ourH1} at $h=0$. We do not see a phase transition in this system as we vary $J$ and $h$. 

The paper is organized as follows. Section \ref{MPSsection} is a review of the MPS ansatz
and contains its generalization to the tree geometry. 
In Section \ref{updatesection}, we review the numerical procedure for unitary updates and give a recipe for applying imaginary time evolution within the MPS ansatz.
In Section \ref{TIsection}, we 
adapt Vidal's iTEBD method for simulating translationally invariant one-dimensional systems to systems with tree geometry.
Section \ref{NRsection} contains our numerical results for the quantum Ising model in a transverse field for translationally invariant systems.
In Section \ref{NRsection1}, we test our method for the infinite line,
and in Section \ref{NRsection2} we present new results for the infinite tree.
We turn to the \textsc{not} 00 model in Section \ref{not00section} and show that our numerics work well for this system even when there is a high ground state degeneracy. In Section \ref{stabilitysection} we investigate the stability of our tree results and conjecture that they may be good approximations to a local description far from the boundary of a large finite tree system.


\section{Matrix Product States} 
\label{MPSsection}

If one's goal is to numerically investigate a system governed by a local Hamiltonian,
it is convenient to find a local description and update rules for the system.
A Matrix Product State description is particularly suited to spin systems
for which the connections do not form any loops. Given a state of this system, we will first 
show how to obtain its MPS description, and then how to utilize this description 
in a numerical method for obtaining the time evolution and approximating the ground state (using imaginary time evolution). We begin with matrix product states on a line (a spin chain),  and then generalize the description to a tree geometry. In \ref{updatesection}, we give a numerical method of updating the MPS description for both real and imaginary time simulations.

\subsection{MPS for a spin chain}

Given a state $\ket{\psi}$ of a chain of $n$ spins
\begin{eqnarray}
  \ket{\psi} &=& \sum_{\dots s_i s_{i+1} \dots}
  c_{\dots,s_i,s_{i+1},\dots}
  \ket{s_1}_1 \dots \ket{s_i}_i \ket{s_{i+1}}_{i+1} \dots \ket{s_n}_n, 
  \label{psiMPS}
\end{eqnarray}
we wish to rewrite the coefficients $c_{s_1,\dots,s_n}$ as a matrix product (see \cite{MPSreview} for a review of MPS)
\begin{eqnarray}
  c_{\dots,s_i,s_{i+1},\dots} &=& 
  \sum_{\dots a b c \dots} \dots \lambda^{(i-1)}_a 
  \Gamma^{(i),s_i}_{a,b} \lambda^{(i)}_b 
  \Gamma^{(i+1),s_{i+1}}_{b,c} \lambda^{(i+1)}_c
  \dots 
  \label{cMPS}
\end{eqnarray}
using $n$ tensors $\Gamma^{(i)}$ and $n-1$ vectors $\lambda^{(i)}$. 
The range of the indices $a,b,\dots$ will be addressed later. 
After decomposing the chain into two subsystems, one can rewrite the state of the whole system
in terms of orthonormal bases of the subsystems. 
$\lambda^{(i)}$ is the vector of Schmidt coefficients 
for the decomposition of the state of the chain onto the subsystems $1\dots i$ and $i+1 \dots n$. 

In order to obtain the $\lambda$'s and the $\Gamma$'s for a given state $\ket{\psi}$, one has to perform the following steps. First, perform the Schmidt decomposition of the chain between sites $i-1$ and $i$ as
\begin{eqnarray}
  \ket{\psi} = \sum_{a=1}^{\chi_{i-1}} \ket{\phi_a}_{1,\dots,i-1} 
  \lambda^{(i-1)}_a \ket{\phi_a}_{i,\dots,n},
\end{eqnarray}
where the states on the left and on the right of the division form
orthonormal bases required to describe the respective subsystems 
of the state $\ket{\psi}$.
The number $\chi_{i-1}$ (the Schmidt number) is the minimum number of terms 
required in this decomposition. 
\begin{figure}
	\begin{center}
	\includegraphics[width=3.3in]{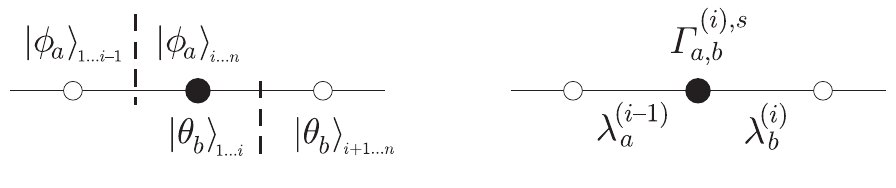}
	\caption{Two successive Schmidt decompositions on a line allow us to find the $\Gamma$
	tensor for the marked site and the two $\lambda$ vectors for the bonds coming out of it.}
	\label{figlineschmidt}
	\end{center}
\end{figure}
The Schmidt decomposition for a split between sites $i$ and $i+1$ gives
\begin{eqnarray}
  \ket{\psi} = \sum_{b=1}^{\chi_{i}} \ket{\theta_b}_{1,\dots,i} 
  \lambda^{(i)}_b \ket{\theta_b}_{i+1,\dots,n}.
  \label{division2}
\end{eqnarray}
These two decompositions (see FIG.\ref{figlineschmidt}) describe the same state, allowing us to combine them to express 
the basis of the subsystem $i,\dots,n$ using the spin at site $i$ and the basis of the subsystem $i+1,\dots,n$ as
\begin{eqnarray}
   \ket{\phi_a}_{i,\dots,n} = \sum_{s=0,1} \sum_{b=1}^{\chi_{i}} \Gamma^{(i),s}_{a,b} 
   \lambda^{(i)}_b \ket{s}_i \ket{\theta_b}_{i+1,\dots,n}, \label{combineSchmidt}
\end{eqnarray}
where we inserted the $\lambda_b^{(i)}$ for convenience. 
This gives us the tensor $\Gamma^{(i)}$. 
It carries an index 
$s$ corresponding to the state $\ket{s}$ of the $i$-th spin, and indices $a$ and $b$, corresponding
to the two consecutive divisions of the system (see FIG.\ref{figlineschmidt}). 
Because $\ket{\phi_a}$ (and $\ket{\theta_b}$) are orthonormal states,
the vectors $\lambda$ and tensors $\Gamma$ obey the following normalization conditions. From \eqref{division2} we have 
\begin{eqnarray}
	\sum_{b=1}^{\chi_i} \lambda_b^{(i)2} = 1,
	\label{normal0}
\end{eqnarray}
while
\eqref{combineSchmidt} implies 
\begin{eqnarray}
	\braket{\phi_{a'}}{\phi_a}_{i,\dots,n} = \sum_{s=0,1} \sum_{b=1}^{\chi_i}  \Gamma^{(i),s *}_{a',b} \lambda_b^{(i)}\Gamma^{(i),s}_{a,b}  \lambda_b^{(i)} = \delta_{a,a'} \, ,
	\label{normal1}
\end{eqnarray}
and 
\begin{eqnarray}
	\braket{\theta_{b'}}{\theta_b}_{1,\dots,i} =
	\sum_{s=0,1} \sum_{a=1}^{\chi_{i-1}} \lambda_a^{(i-1)} \Gamma^{(i),s *}_{a,b'}
	\lambda_a^{(i-1)} \Gamma^{(i),s}_{a,b} = \delta_{b,b'} \,  .
	\label{normal2}
\end{eqnarray}


\subsection{MPS on Trees} 
\label{MPStreeSubsection}

Matrix Product States are natural not just on chains, but also on trees,
because these can also be split into two subsystems by cutting a single bond, 
allowing for the Schmidt-decomposition interpretation
as described in the previous section.
The Matrix Product State description of a state of a spin system on a tree, 
i.e. such that the bonds do not form loops, is a generalization of the above procedure. 
Tree-tensor-network descriptions such as ours have been previously described in \cite{ShiTrees}.

Specifically, for the Bethe lattice with 3 neighbors per spin,
we introduce a vector $\lambda^{(k)}_{a_k}$ for each bond $k$ and a four-index (one for spin, three for bonds) tensor $\Gamma^{(i),s_i}_{a_k,a_l,a_m}$ for each site $i$. We can then rewrite the state $\ket{\psi}$ analogously to \eqref{psiMPS},\eqref{cMPS} as
\begin{eqnarray}
	\ket{\psi} = 
		\Bigg(
			\prod_{k\in\textrm{bonds}} 
			\sum_{a_k = 1}^{\chi_k}
			\lambda^{(k)}_{a_k} 
		\Bigg)
		\Bigg(
			\prod_{i\in \textrm{sites}} 
			\sum_{s_i} 
		\Gamma^{(i),s_i}_{a_l, a_m, a_n} 
		\Bigg)
	\ket{\dots}\ket{s_i}\ket{\dots},
\end{eqnarray}
where $a_l,a_m,a_n$ are indices corresponding to the three bonds $l, m$ and $n$ coming out of site $i$. Each index $a_l$ appears in two $\Gamma$ tensors and one $\lambda$ vector. To obtain this description, one needs to perform a Schmidt decomposition across each bond. This produces 
the vectors $\lambda^{(l)}$.
\begin{figure}
	\begin{center}
	\includegraphics[width=3.5in]{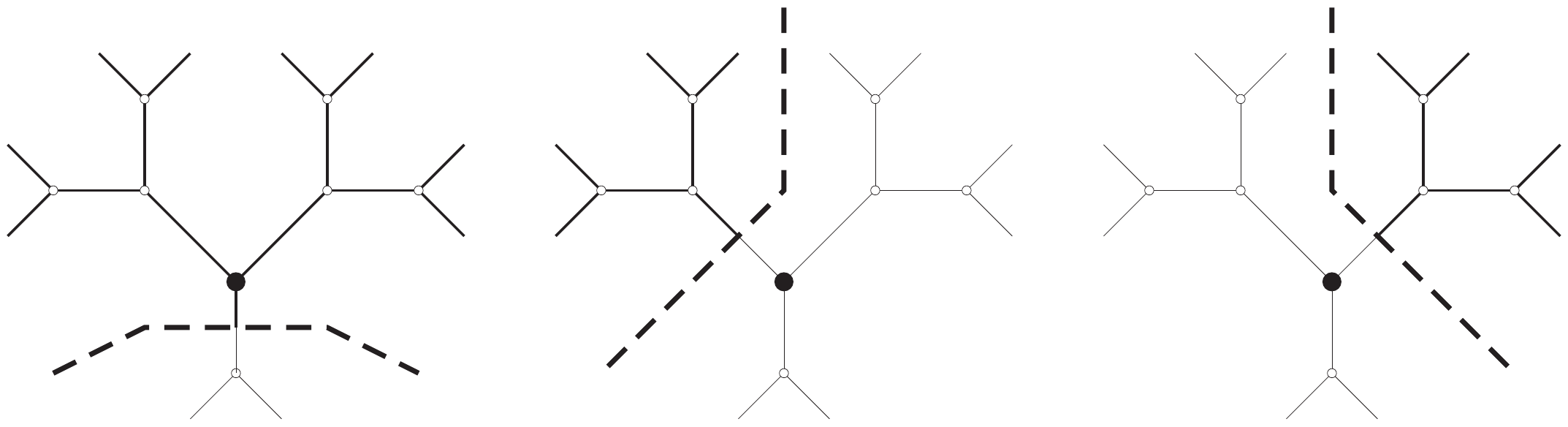}
	\caption{The three Schmidt decompositions on a tree required to obtain the $\Gamma$ tensor
	for the marked site and the three $\lambda$ vectors for the bonds emanating from it.}
	\label{figtreeschmidt}
	\end{center}
\end{figure}
To obtain the tensor $\Gamma^{(i)}$ for site $i$, 
one needs to combine the three decompositions corresponding to the bonds of site $i$
as depicted in Fig.\ref{figtreeschmidt}.
Analogously to \eqref{combineSchmidt}, expressing the orthonormal basis
for the first subsystem marked in Fig.\ref{figtreeschmidt} in terms of the state of the spin $\ket{s_i}$ and the orthonormal bases for the latter two subsystems in Fig.\ref{figtreeschmidt}, one obtains the tensor $\Gamma^{(i),s}_{a_l,a_m,a_n}$ for site $i$. 

The normalization conditions for a MPS description of a state on a tree are analogous to \eqref{normal0}-\eqref{normal2}.
We have
\begin{eqnarray}
	\sum_{a_k} \lambda^{(k)2}_{a_k} = 1, 
		\label{normalT0}
\end{eqnarray}
\begin{eqnarray}
	\sum_{s=0,1} \sum_{a_k = 1}^{\chi_k} \sum_{a_l = 1}^{\chi_l} \Gamma^{(i),s *}_{a_k,a_l,a_{m'}} \lambda^{(k)2}_{a_k}\lambda^{(l)2}_{a_l} \Gamma^{(i),s}_{a_k,a_l,a_m} = \delta_{a_m,a_{m'}} \, , 
	\label{normalT1}
\end{eqnarray}
and two other variations of \eqref{normalT1} with $k,l$ and $m$ interchanged.


\section{Simulating Quantum Systems with MPS}
\label{updatesection}

We choose to first describe the numerical procedures for a chain of spins. Then, at the end of the respective subsections, we note how to generalize these to tree geometry.


\subsection{Unitary Update Rules}

The strength of the MPS description of the state lies in the efficient 
application of local unitary update rules such as $U = e^{-i A \Delta t}$ (where $A$ is an operator acting only on a few qubits). 
First, we describe the numerical procedure in some detail, and then, in the next Section,  discuss how to modify the procedure to also implement imaginary time evolution.

Given a state $\ket{\psi}$ as a Matrix Product State, we want to know what happens after 
an application of a local unitary. In particular, for a 1-local $U$ acting on the $i$-th spin, 
it suffices to update the local tensor
\begin{eqnarray}
	\Gamma^{(i),s}_{a,b} 
	\buildrel{U}\over\longrightarrow U^{s}_{s'} \Gamma^{(i),s'}_{a,b}.
\end{eqnarray} 
The update rule for an application of a 2-local unitary $V$ acting on neighboring spins $i$ and $i+1$,
requires several steps. First, using
a larger tensor 
\begin{eqnarray}
	\Theta^{s,t}_{a,c} = \lambda^{(i-1)}_a 
	  \sum_b \left( \Gamma^{(i),s}_{a,b}
	  \lambda^{(i)}_b \Gamma^{(i+1),t}_{b,c} \right) \lambda^{(i+1)}_c,
	  \label{Thetatensor}
\end{eqnarray}
we rewrite the state $\ket{\psi}$ as
\begin{eqnarray}
  	\ket{\psi} &=& \sum_{a,c} \sum_{s,t} \Theta^{s,t}_{a,c} \ket{\phi_a}_{1\dots i-1}
	  \ket{s}_{i} \ket{t}_{i+1} \ket{\phi_c}_{i+1 \dots n}.
	  \label{Thetatensor2}
\end{eqnarray}
After the application of $V$, the tensor $\Theta$ in the description of $\ket{\psi}$ 
changes as
\begin{eqnarray}
	\Theta^{s,t}_{a,c} \buildrel{V}\over\longrightarrow 
	\sum_{s' t'} V^{s,t}_{s',t'} \Theta^{s',t'}_{a,c}.
	\label{Thetaupdate}
\end{eqnarray}
One now needs to decompose the updated tensor $\Theta$ to obtain the updated 
tensors $\Gamma^{(i)}$, $\Gamma^{(i+1)}$ and the vector $\lambda^{(i)}$.
We use the indices $a,s$ and $c,t$ of $\Theta$
to introduce combined indices $(as)$ and $(ct)$ and form a matrix $T_{(as),(ct)}$ 
with dimensions $2\chi_{i-1} \times 2\chi_{i+1}$ as
\begin{eqnarray}
	T_{(as),(ct)} =  \Theta_{a,c}^{s,t}.
	\label{Tmatrix1}
\end{eqnarray}
Using the singular value decomposition (SVD), this matrix can be decomposed into $T = Q \Lambda W$, where
$Q$ and $W$ are unitary and $\Lambda$ is a diagonal matrix. 
In terms of matrix elements, this reads
\begin{eqnarray}
	T_{(as),(ct)} = 
  \sum_b Q_{(as),b} 
  D_{b,b} W_{b,(ct)}.
  \label{Tmatrix}
\end{eqnarray}
The diagonal matrix $D=\textrm{diag}(\lambda^{(i)})$
gives us the updated Schmidt vector 
$\lambda^{(i)}$. 
The updated tensors $\Gamma^{(i)}$ and $\Gamma^{(i+1)}$ can be obtained from the 
matrices $Q,W$ and the definition of $\Theta$ \eqref{Thetatensor} 
using the old vectors $\lambda^{(i-1)}$ and $\lambda^{(i+1)}$ which do 
not change with the application of the local unitary $V$. 
After these update procedures, the conditions \eqref{normal0}-\eqref{normal2}
are maintained.

The usefulness/succintness of this description depends crucially on the amount of entanglement 
across the bipartite divisions of the system as measured by the Schmidt numbers $\chi_i$.
To exactly describe a general quantum state $\ket{\psi}$ of a chain of $n$ spins, the Schmidt 
number for the split through the middle of the chain is necessarily $\chi_{n/2} = 2^{n/2}$.
Suppose we start our numerical simulation in a state that is exactly described by a MPS with only low $\chi_i$'s.
The update step described above involves an interaction
of two sites, and thus could generate more entanglement across the $i,i+1$ division.
After the update, the index $b$ in $\lambda^{(i)}_b$ 
would need to run from $1$ to $2\chi_i$ to keep 
the description exact (unless $\chi_i$ already is at its maximum required value 
$\chi_i=2^{\textrm{min}\{i,n-i\}}$).
This makes the number of parameters in the MPS description grow exponentially 
with the number of update steps.

So far, this description and update rules have been exact. 
Let us now make the description an approximate one (use a block-decimation step) instead. 
First, introduce the parameter $\chi$, which is the maximum number of Schmidt terms we  keep after each update step.
If the amount of entanglement in the system is low, the Schmidt coefficients 
$\lambda^{(i)}_b$ decrease rapidly with $b$ (We always take the elements of $\lambda$ sorted in decreasing order).
A MPS ansatz with restricted $\chi_i = \chi$ will hopefully be a good approximation to 
the exact state $\ket{\psi}$. 
However, we also need to keep the restricted $\chi$ throughout the simulation.
After a two-local unitary update step, the vector $\lambda^{(i)}$ can have $2\chi$ entries. However, if the $b>\chi$ entries in $\lambda^{(i)}_b$ after the update are small,
we are justified to truncate $\lambda^{(i)}$ to have only $\chi$ entries and multiply it 
by a number so that it satisfies \eqref{normal0}.
We also truncate the $\Gamma$ tensors so that they keep dimensions $2\times\chi\times\chi$. The normalization condition \eqref{normal2} for $\Gamma^{(i)}$ will be still satisfied exactly, while the error in the normalization condition \eqref{normal1} will be small. This normalization error can be corrected as discussed in the next section.
This procedure keeps us within the MPS ansatz with restricted $\chi$.

The procedure described above allows us to efficiently approximately implement local unitary evolution. To simulate time evolution 
\begin{eqnarray}
	\ket{\psi(t)} = e^{-iHt}\ket{\psi(0)}, \label{realtime}
\end{eqnarray}
with a local Hamiltonian like \eqref{ourH1}, we first divide the time $t$ into small slices $\Delta t$ and split the Hamiltonian into two groups of commuting terms $H_k^{(x)}$ and $H_m^{(z)}$. Each time evolution step $e^{-iH\Delta t}$ can then be implemented as a product of local unitaries using the second order Trotter-Suzuki formula
\begin{eqnarray}
	U_2 = \left(\prod_{k} e^{-i H^{(x)}_k \frac{\Delta t}{2}}\right)
		\left(\prod_{m} e^{-i H^{(z)}_m \Delta t}\right)
		\left(\prod_{k} e^{-i H^{(x)}_k \frac{\Delta t}{2}}\right).
		\label{trotter}
\end{eqnarray}
The application of the product of the local unitaries within each group can be done almost in parallel (in two steps, as described in Section \ref{TIsection}), as they commute with each other.

These update rules allow us to efficiently approximately simulate the real 
time evolution \eqref{realtime}
with a local Hamiltonian $H$ 
for a state $\ket{\psi}$ within the MPS ansatz with parameter $\chi$. 
The number of parameters in this MPS description with restricted $\chi$  
is then $n (2\chi^2)$ for the tensors $\Gamma^{(i)}$ and $(n-1) \chi$ for the vectors 
$\lambda^{(i)}$.
The simulation cost of each local update step scales like $O(\chi^3)$, 
coming from the SVD decomposition of the matrix $\Theta$.
For a system of $n$ spins, we thus need to store $O(2n\chi^2+n\chi)$ numbers
and each update will take $O(n\chi^3)$ steps.

The update procedure generalizes to tree geometry by taking the tensors $\Gamma$ with dimensions $2\times\chi\times\chi\times \chi$ as in Section \ref{MPStreeSubsection}. 
For a local update (on two neighboring spins $i$ and $i+1$ with bonds labeled by $l,m,n$ and $n , o, p$) we rewrite the state $\ket{\psi}$ analogously to \eqref{Thetatensor2} as
\begin{eqnarray}
  	\ket{\psi} &=& \sum_{a_k,a_l,a_o,a_p} \sum_{s,t} \Theta^{s,t}_{(a_k a_l),(a_o a_p)} 
	\ket{\phi_{a_k}} \ket{\phi_{a_l}}
	  \ket{s}_{i} \ket{t}_{i+1} \ket{\phi_{a_o}}\ket{\phi_{a_p}}.
\end{eqnarray}
using the tensor 
\begin{eqnarray}
	\Theta^{s,t}_{(a_k a_l),(a_o a_p)} = \lambda^{(k)}_{a_k} \lambda^{(l)}_{a_l} 
	  \sum_{a_m} \left( \Gamma^{(A),s}_{a_k,a_l,a_m}
	  \lambda^{(m)}_{a_m} \Gamma^{(i+1),t}_{a_m,a_o,a_p} \right) \lambda^{(o)}_{a_o} \lambda^{(p)}_{a_p},
\end{eqnarray}
with combined indices $(a_k a_l)$ and $(a_o a_p)$.
One then needs to update the tensor $\Theta$ as described above \eqref{Thetaupdate}-\eqref{Tmatrix}. The decomposition procedure to get the updated vector $\lambda^{(m)}$ and the new tensors $\Gamma^{(i)}$ and $\Gamma^{(i+1)}$ now requires 
$O(\chi^6)$ computational steps. The cost of a simulation on $n$ spins thus scales like $O(n\chi^6)$.


\subsection{Imaginary Time Evolution}
\label{imaginarysection}

Using the MPS ansatz, we can also use imaginary time evolution with $e^{-Ht}$ instead of \eqref{realtime} to look for the ground state of systems governed by local Hamiltonians. One needs to replace each unitary term
$e^{-i A \Delta t}$ 
in the Trotter expansion \eqref{trotter} of the time evolution with
$e^{- A \Delta t}$ followed by a normalization procedure. However, the usual normalization procedure for imaginary time evolution (multiplying the state by a number to keep $\braket{\psi}{\psi}=1$) is now not enough to satisfy the MPS normalization conditions \eqref{normal0}-\eqref{normal2} for the tensors $\Gamma$ and vectors $\lambda$ we use to describe the state $\ket{\psi}$.

The unitarity of the real time evolution automatically implied
that the normalization conditions \eqref{normal0},\eqref{normal2} were satisfied after an
exact unitary update. While there already was an error in \eqref{normal1} introduced by the truncation of the $\chi+1 \dots 2\chi$ entries in $\Gamma^{(i)}$, the non-unitarity of imaginary time evolution update steps introduces further normalization errors. It is thus important to properly normalize the state after every application of terms like $e^{-A\Delta t}$ to keep it within the MPS ansatz. 

In \cite{Vidal1Dinfinite}, Vidal dealt with this problem by taking progressively shorter and shorter steps $\Delta t$ during the imaginary time evolution. This procedure results in a properly normalized state only at the end of the evolution, after the time step decreases to zero (and not necessarily during the evolution). 
We propose a different scheme in which we follow each local update $e^{-A \Delta t}$ by a normalization procedure (based on Vidal's observation) to bring the state back to the MPS ansatz at all times. 
The simulation we run (evolution for time $t$) thus consists of many short time step updates $e^{-H \Delta t}$, each of which is implemented using a Trotter expansion as a product of local updates $e^{-A \Delta t}$. Each of these local updates is followed by our normalization procedure.
 
We now describe the iterative normalization procedure in detail for the case of an infinite chain, where it can be applied efficiently, as the description of the state $\ket{\psi}$ requires only two different tensors $\Gamma$ (see Section \ref{TIline}). One needs to apply the following steps over and over, until the normalization conditions are met with chosen accuracy. 

First, for each nearest neighbor pair $i,i+1$ with even $i$, combine the MPS description of these two spins \eqref{Thetatensor}-\eqref{Thetatensor2}, forming the matrix $T$ \eqref{Tmatrix1}. Do a SVD decomposition of $T$ \eqref{Tmatrix} to obtain a new vector $\lambda^{(i)}$. The decomposition does not increase the number of nonzero elements of $\lambda^{(i)}$, as the rank of the $2\chi\times2\chi$ matrix $T$ \eqref{Tmatrix} was only $\chi$ (coming from \eqref{Thetatensor}).
We thus take only the first $\chi$ values of $\lambda^{(i)}$
and rescale the vector to obey $\sum_{a=1}^{\chi} \lambda^{(i)2}_a = 1$.
using this new $\lambda^{(i)}$, we obtain tensors $\Gamma^{(i)}$, $\Gamma^{(i+1)}$ from \eqref{Tmatrix}, and truncate them to have dimensions $\chi\times\chi\times2$.  
Second, we repeat the previous steps for all nearest neighbor pairs of spins $i,i+1$ with $i$ odd. 

We observe that repeating the above steps over and over results in exponential decrease in the error in the normalization of the $\Gamma$ tensors. We note though, that the rate of decrease in normalization errors becomes much slower near the phase transition for the transverse field Ising model on an infinite line (see Section \ref{NRsection1}).

In practice, we apply this normalization procedure by using the same subroutine for the local updates $e^{-A\Delta t}$, except that we skip the step 
\eqref{Thetaupdate}, which is equivalent to applying the local update with $\Delta t=0$.
The normalization procedure is thus equivalent to evolving the state repeatedly with zero time step (composing two tensors $\Gamma$ and decomposing them again) and imposing the normalization condition on the vectors $\lambda$. Note though, following from the definition of the SVD, that each decomposition assures us that one of the conditions \eqref{normal1},\eqref{normal2} is retained exactly for the updated tensors $\Gamma$. The errors in the other normalization condition for the $\Gamma$ tensors 
are decreased in each iteration step.

The numerical update rules for a system with tree geometry are a simple analogue of the update rules for MPS on spin chains. Every interaction couples two sites, 
with tensors $\Gamma^{(A),s}_{a,b,c}$ and $\Gamma^{(B),t}_{c,d,e}$, with the three lower indices corresponding to the bonds emanating from the sites. 
One only needs to reshape the tensors into $\Gamma^{(A),s}_{(ab),c}$ 
and $^{(B),t}_{c,(de)}$ and proceed as described in \eqref{Thetatensor}
and below.


\section{MPS and Translationally Invariant Systems} 
\label{TIsection}

\subsection{An Infinite Line} 
\label{TIline}

For systems with translational symmetry such as an infinite line 
all the sites are equivalent. 
We assume that the ground state is translationally invariant, and
furthermore pick the tensors $\Gamma^{(i)}$ and vectors $\lambda^{(i)}$ 
to be site independent.
For fixed $\chi$ the number of complex parameters in the translationally 
invariant MPS ansatz on the infinite line scales as $2\chi^2$. 

\begin{figure}
	\begin{center}
	\includegraphics[width=2.6in]{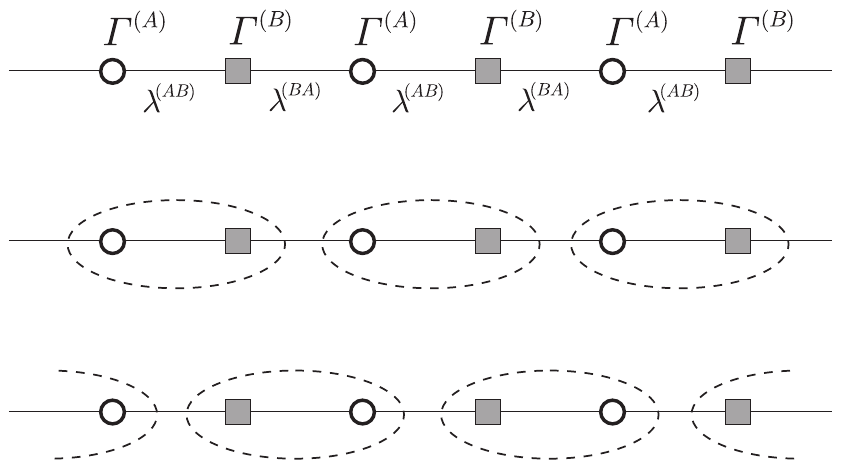}
	\caption{The parametrization and update rules for the infinite line.}
	\label{figlineAB}
	\end{center}
\end{figure}

When using imaginary time evolution to look for the ground state of this system, within this ansatz, 
it is technically hard to keep the translational symmetry and the normalization conditions after each update. 
Numerical instabilities plagued our efforts to impose the symmetry in the 
procedures described above.
In \cite{Vidal1Dinfinite}, Vidal devised a method to deal with this problem. Let us
break the translational symmetry of the ansatz by labeling the sites $A$ and $B$ as in FIG.\ref{figlineAB}.
This doubles the number of parameters in the ansatz.
The state update now proceeds in two steps. Let the site pairs $AB$ 
interact and update the tensors $\Gamma^{(A)}$, $\Gamma^{(B)}$ and the vector $\lambda^{(AB)}$. 
Then let the neighbor pairs $BA$ interact, after which we update the tensors 
$\Gamma^{(B)}$, $\Gamma^{(A)}$ and the vector $\lambda^{(BA)}$.
What we observe is that after many state updates the elements of the resulting $\Gamma^{(A)}$ and the $\Gamma^{(B)}$ tensors differ at a level which is way below our numerical accuracy (governed by the normalization errors)
and we are indeed obtaining a translationally invariant description of the system.

One of the systems easily investigated with this method (iTEBD) is the Ising model in a transverse field \eqref{ourH} on an infinite line.
Vidal's numerical results for the real time evolution and imaginary time 
evolution \cite{Vidal1Dinfinite} of this system show remarkable agreement 
with the exact solution.
We take a step further and also numerically obtain the critical exponents for this system.
Further details can be found in Section \ref{NRsection}, where we compare these
results for the infinite line to the results we obtain for the Ising model
in transverse field on the Bethe lattice.


\subsection{An Infinite Tree} 
\label{infitreesection}

\begin{figure}[h]
	\begin{center}
	\includegraphics[width=2.5in]{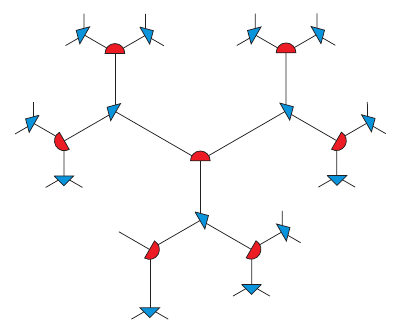}
	\caption{The two-layer, directed labeling of the tree.}
	\label{figtreeAB}
	\end{center}
\end{figure}

For the infinite Bethe lattice, 
our approach is a modification of the above procedure introduced by Vidal. 
In order to avoid the numerical instabilities associated with imposing site-independent $\Gamma$ and $\lambda$ after the update steps, 
we break the translational symmetry by labeling the ``layers'' of the tree $A$ and $B$ 
(denoted by half-circles and triangles), as in FIG.\ref{figtreeAB}. 

\begin{figure}[h]
	\begin{center}
	\includegraphics[width=5in]{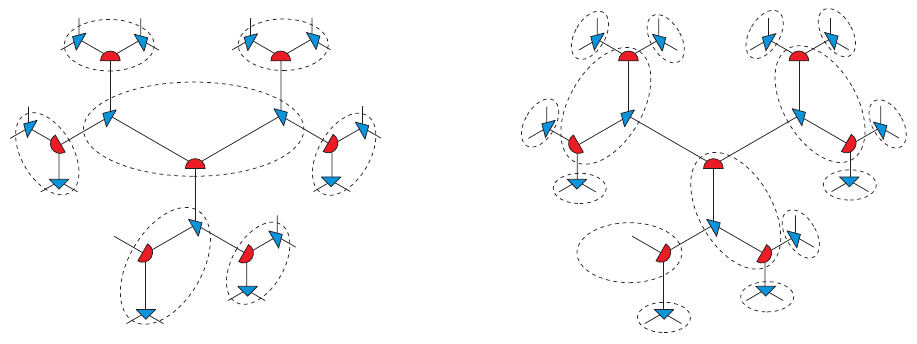}
	\caption{The two-step interactions for the infinite tree.}
	\label{figtreeABinteract}
	\end{center}
\end{figure}

The Bethe lattice is also symmetric under the 
permutation of directions. Tensors $\Gamma$ with full directional symmetry obey
$\Gamma_{a,b,c} = \Gamma_{b,c,a}=\Gamma_{c,a,b}=\Gamma_{c,b,a}=\Gamma_{b,a,c}=\Gamma_{a,c,b}$.
However, for the purpose of simple organization of interactions, we will also
partially break this symmetry by consistently labeling an `inward' bond for each node,
as denoted by the flat sides of the semi-circles and the longer edges of the triangles in Fig.\ref{figtreeAB}.
This makes the first of the three indices of $\Gamma_{a,b,c}$ special. 
However, we keep the residual symmetry $\Gamma_{a,b,c}=\Gamma_{a,c,b}$.
This we can enforce by interacting 
a spin with both of the spins from the next layer at the same time.
The update procedure for the interaction between the spins now splits into two steps,
interacting the layers in the $AB$ order first, 
and then in the $BA$ order as in Fig.\ref{figtreeABinteract}.
Similarly to what we discovered for the line, the differences 
in the elements of the final $\Gamma^{(A)}$ and $\Gamma^{(B)}$ are
well below the numerical accuracy of our procedure.

The scaling of this procedure is more demanding than the $O(\chi^3)$ simulation for a line.
The number of entries in the matrix $\Theta$ used in each update step 
is $2\chi^2 \times 4\chi^4$, therefore the SVD decomposition requires  $O(\chi^8)$
steps. The scaling of our numerical method is thus $O(\chi^8)$ for each update step.


\subsection{Expectation Values}
A nice property of the MPS state description is that it allows efficient computation of expectation values of local operators. 
First, for a translationally invariant system on a line (with only one tensor $\Gamma$ and one vector $\lambda$), we have
for an operator $O^{(i)}$ acting only on the $i$-th spin 
\begin{eqnarray}
	\bra{\psi}O^{(i)} \ket{\psi} = 
	\sum_{s_i,s_i'=0,1} O_{s_i,s_i'}^{(i)}
	\sum_{a=1}^{\chi} 
	\sum_{b=1}^{\chi} 
	(\lambda_{a}
	\Gamma^{s_i' *}_{a,b}
	\lambda_{b} )
	(\lambda_{a}
	\Gamma^{s_i}_{a,b} 
	\lambda_{b} )
	,
	\label{expect1}
\end{eqnarray}
where $O^{(i)}_{s_i,s_i'} = \bras{s_i'}O^{(i)}\kets{s_i}$.
Similarly, for the expectation values of $O^{(i)}O^{(j)}$ (assuming $j>i$), 
\begin{eqnarray}
	\bra{\psi}O^{(i)}O^{(j)} \ket{\psi} 
	&=& 
	\sum_{s_i,s'_i,\dots,s_j,s'_j}
	  \sum_{a,e} \sum_{b,b', \dots} 
	 O_{s_i,s_i'}^{(i)}O_{s_j, s_j'}^{(j)} \label{twoexpect} \\
	&&\times \, \big(\lambda_{a}
			\Gamma^{s_i'*}_{a,b'} 
			\lambda_{b'}
			\Gamma^{s_{i+1}*}_{b',c'} 
			\lambda_{c'}
			\cdots
			\lambda_{d'}
			\Gamma^{s_j'*}_{d',e} 
			\lambda_{e}\big) \nonumber\\
	 &&\times \, 
		\big(\lambda_{a}
			\Gamma^{s_i}_{a,b} 
			\lambda_{b}
			\Gamma^{s_{i+1}}_{b,c} 
			\lambda_{c}
			\cdots
			\lambda_{d}
			\Gamma^{s_j}_{d,e} 
			\lambda_{e}\big). 
	\nonumber
\end{eqnarray}
Defining a $\chi^2 \times \chi^2$ matrix $B$ (where one should think of $(bb')$ as one combined index ranging from $1$ to $\chi^2$) as 
\begin{eqnarray}
	B_{(bb'), (cc')} = 
	\sum_{s} 
			\Gamma^{s}_{b,c}
			\Gamma^{s *}_{b',c'}
			\lambda_{c}
			\lambda_{c'},
\end{eqnarray}
and
vectors $v$ and $w$ with elements again denoted by a combined index $(bb')=1\dots\chi^2$ as
\begin{eqnarray}
	v_{(bb')} &=& 
		\sum_{s_i,s_i'} O_{s_i,s_i'}^{(i)}
	\sum_{a} 
			(\lambda_{a})^2
			\Gamma^{s_i}_{a,b} 
			\Gamma^{s_i' *}_{a,b'} 
			\lambda_{b}
			\lambda_{b'}
			, \\
	w_{(dd')} &=& 
		\sum_{s_j,s_j'} O_{s_j,s_j'}^{(j)}
	\sum_{e} 
			\Gamma^{s_j}_{d,e} 
			\Gamma^{s_j' *}_{d',e} 
			(\lambda_{e})^2
			,
\end{eqnarray}
we can rewrite \eqref{twoexpect} as
\begin{eqnarray}
	\bra{\psi}O^{(i)}O^{(j)} \ket{\psi} = v^T \underbrace{B B \cdots B}_{j-i-1} w,
	\label{Bcorrfunction}
\end{eqnarray}

There is a relationship between the eigenvalues of the matrix $B$ and the correlation function $\corrfO$. One of the eigenvalues of $B$ is $\mu_1=1$, with the corresponding right eigenvector  
\begin{eqnarray}
	\beta^{(1R)}_{(cc')} =   
	\sum_{s} 
	\sum_{c}
			\Gamma^{s}_{b,c}
			\Gamma^{s *}_{b',c}
			(\lambda_{c})^2,
\end{eqnarray}
and left eigenvector
\begin{eqnarray}
	\beta^{(1L)}_{(bb')} =   
		\delta_{b,b'} \lambda_{b}^2,
\end{eqnarray}
which can be verified using the normalization conditions \eqref{normal1} and \eqref{normal2}.
We numerically observe that $\mu_1=1$ is also the largest eigenvalue.  
(Note that $|\mu_k|>1$ would result in correlations unphysically growing with distance.)
Denote the second largest eigenvalue of $B$ as $\mu_2$. 
Using the eigenvectors of $B$, we can express $B^{j-i-1}$ in \eqref{Bcorrfunction}, as
\begin{eqnarray}
	B^{j-i-1} = \beta^{(1L)} \beta^{(1R)T} + \mu_2^{j-i-1} \beta^{(2L)} \beta^{(2R)T} + \dots.
	\label{eigexpand}
\end{eqnarray}
When computing the correlation function, 
the term that gets subtracted exactly cancels the leading term involving $\mu_1=1$. 
Therefore, if $|\mu_2|$ is less than 1, \eqref{eigexpand} implies
\begin{eqnarray}
	\corrfO \propto \mu_2^{|j-i|}. \label{secondeig}
\end{eqnarray} 
The correlation function necessarily falls of exponentially in this case,
and the correlation length $\xi$ is related to $\mu_2$ as $\xi = -1/\ln \mu_2$.

The computation of expectation values for a MPS state on a system with a tree geometry can be again done efficiently. For single-site operators $O^{(i)}$, the formula is an analogue of \eqref{expect1} with three $\lambda$ vectors for each $\Gamma$ tensors which now have three lower indices. For two-site operators, the terms in \eqref{Bcorrfunction} now become
\begin{eqnarray}
	B_{(cc'), (dd')} &=& 
	\sum_{s} \sum_{e}
			\Gamma^{s}_{c,e,d}
			\Gamma^{s *}_{c',e,d'}
			(\lambda_{e})^2
			\lambda_{d}
			\lambda_{d'},\\
	v_{(cc')} &=& 
		\sum_{s_i,s_i'} O_{s_i,s_i'}^{(i)}
	\sum_{a,b} 
			(\lambda_{a})^2 (\lambda_{b})^2
			\Gamma^{s_i}_{a,b,c} 
			\Gamma^{s_i' *}_{a,b,c'} 
			\lambda_{c}
			\lambda_{c'}
			, \\
	w_{(dd')} &=& 
		\sum_{s_j,s_j'} O_{s_j,s_j'}^{(j)}
	\sum_{e,f} 
			\Gamma^{s_j}_{d,e,f} 
			\Gamma^{s_j' *}_{d',e,f} 
			(\lambda_{e})^2
			(\lambda_{f})^2
			.
\end{eqnarray}
The correlation length is again related to the second eigenvalue of the $B$ matrix as
in \eqref{secondeig}.


\section{Quantum Transverse Field Ising Model} 
\label{NRsection}

Our goal is to investigate the phase transition for the Ising model in
transverse magnetic field \eqref{ourH1} on the infinite line and on the Bethe lattice. 
We choose to parametrize the Hamiltonian as  
\begin{eqnarray}
	H = \frac{s}{2} \sum_{\langle i,j\rangle} \left(1-\sigma_z^{i} \sigma_z^{j}\right) 
		+ \frac{b(1-s)}{2} \sum_i \left(1-\sigma_x^{i}\right), \label{ourH}
\end{eqnarray}
where $0\leq s\leq 1$
and $b$ is the number of bonds for each site ($b=2$ for the line, $b=3$ for the tree).
We will investigate
the ground state properties of \eqref{ourH} as we vary $s$. 
The point $s=0$ corresponds
to a spin system in transverse magnetic field, while $s=1$ corresponds
to a purely ferromagnetic interaction between the spins.


\subsection{The Infinite Line.}
\label{NRsection1}
We present the results for the case of an infinite line and compare
them to exact results obtained via fermionization 
(see e.g. \cite{Sachdevbook}, Ch.4).
Vidal has shown \cite{Vidal1Dinfinite} that imaginary time evolution within the MPS ansatz is capable of providing a very accurate approximation for the ground state energy and correlation function. We show that even using $\chi$ smaller than used in \cite{Vidal1Dinfinite}, we obtain the essential information about the nature of the phase 
transition in the infinite one-dimensional system. We also 
obtain the critical exponents for the magnetization and the correlation length.

\begin{figure}
	\begin{center}
	\includegraphics[width=4.5in]{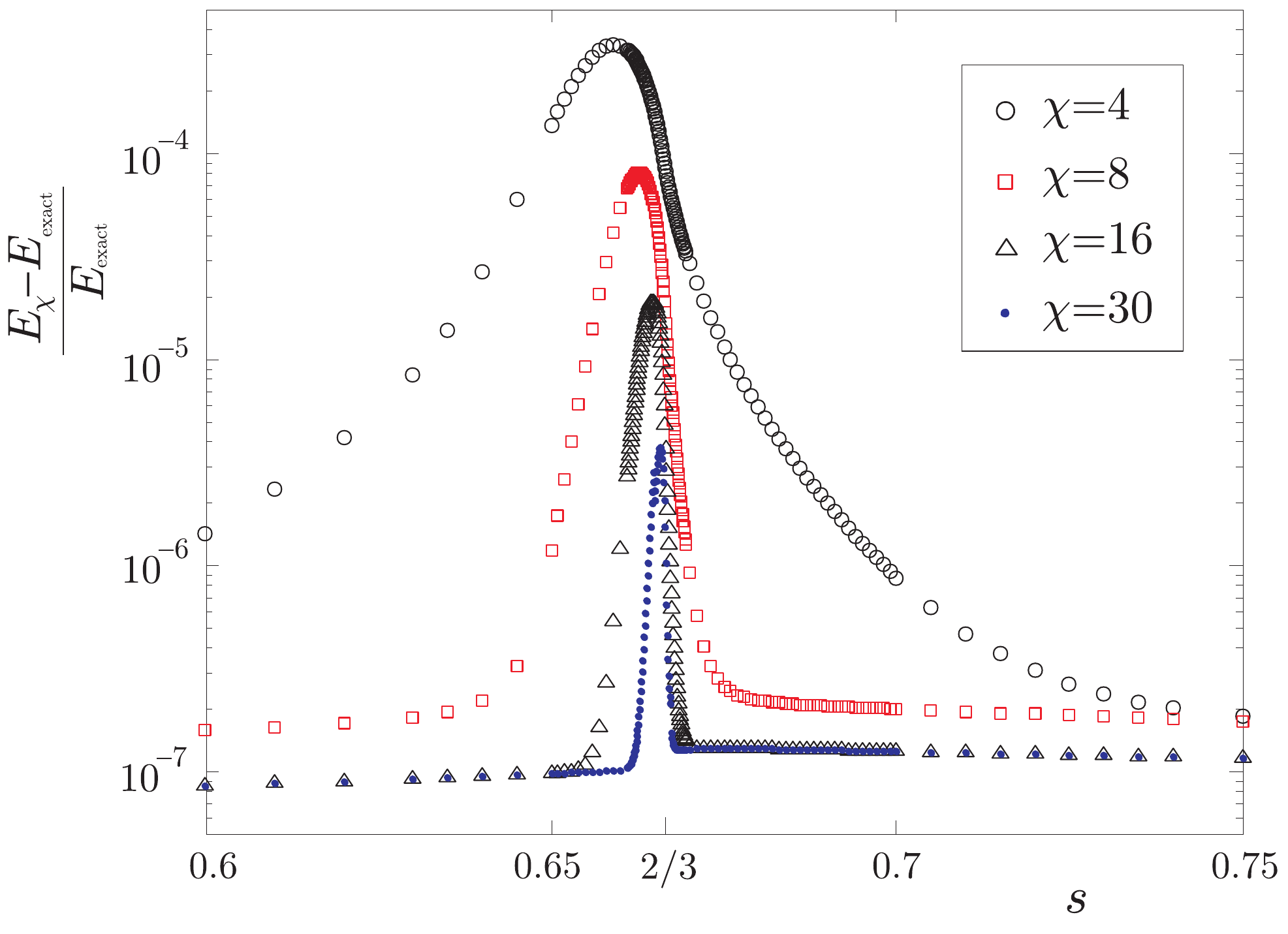}
	\caption{Transverse Ising model on an infinite line. Fractional difference of the ground state energy obtained using MPS and the exact ground state, near the phase transition at $s_L=\frac{2}{3}$. The energy scale is logarithmic.
	}
	\label{lineEd}
	\end{center}
\end{figure}

\begin{figure}
	\begin{center}
	\includegraphics[width=6in]{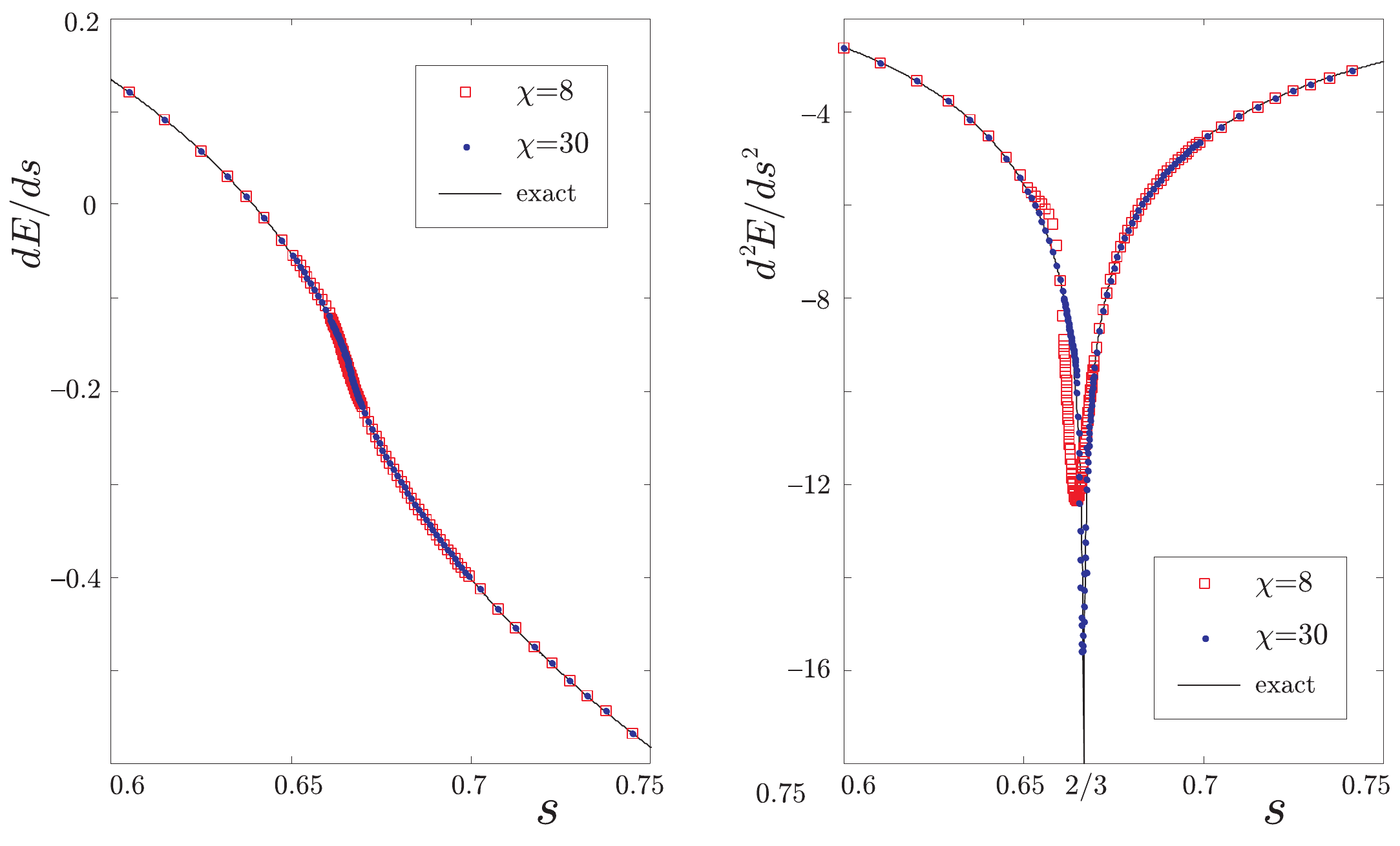}
	\caption{Transverse Ising model on an infinite line. The first and second derivative with respect to $s$ of the
	ground state energy obtained via MPS compared with the exact result.
	}
	\label{lineE12}
	\end{center}
\end{figure}

In FIG.\ref{lineEd}, we show the how the ground state energy obtained using imaginary time evolution with MPS converges to the exact energy as $\chi$ increases.
The exact solution for a line has a second order phase transition 
at the critical value of $s$, $s_L=\frac{2}{3}$, and the ground state energy 
and its first derivative are continuous,
while the second derivative diverges at $s=s_L$.
We plot the first and second derivative of $E$ with respect to $s$ obtained numerically and compare them to the exact values in FIG. \ref{lineE12},
observing the expected behavior already for low $\chi$.

\begin{figure}
	\begin{center}
	\includegraphics[width=3.5in]{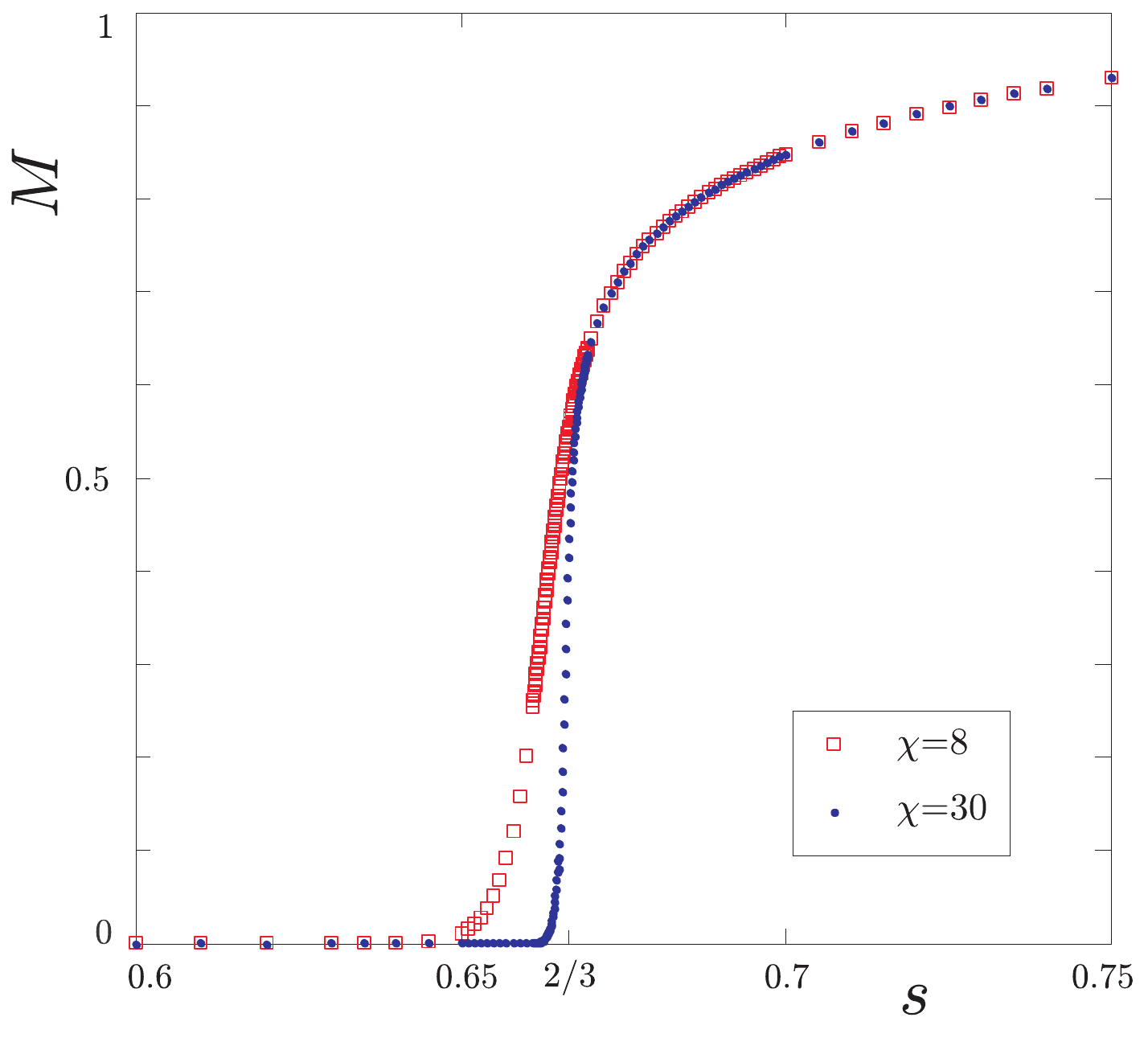}
	\caption{Transverse Ising model on an infinite line. Magnetization obtained using MPS vs $s$, with $B_z=10^{-8}$.
	}
	\label{lineM}
	\end{center}
\end{figure}
The derivative of the exact magnetization $M = \langle \sigma_z \rangle$ is 
discontinuous at $s_L$, with the magnetization starting to rise steeply
from zero as 
\begin{eqnarray}
	M \propto (x-x_L)^{\beta},
	\label{Mexponentline}
\end{eqnarray}
with the critical exponent $\beta_L = \frac{1}{8}$. Here $x$ is the ratio of the ferromagnetic interaction strength
to the transverse field strength in \eqref{ourH} and is
\begin{eqnarray}
	x = \frac{s}{2(1-s)},
\end{eqnarray}
with the value $x=x_L=1$ at the phase transition ($s_L=\frac{2}{3}$). 
We plot the magnetization obtained with our method 
in FIG.\ref{lineM}. To obtain the magnetization depicted in the plot, we used a small symmetry breaking longitudinal field with magnitude $B_z = 10^{-8}$.
\begin{figure}
	\begin{center}
	\includegraphics[width=4in]{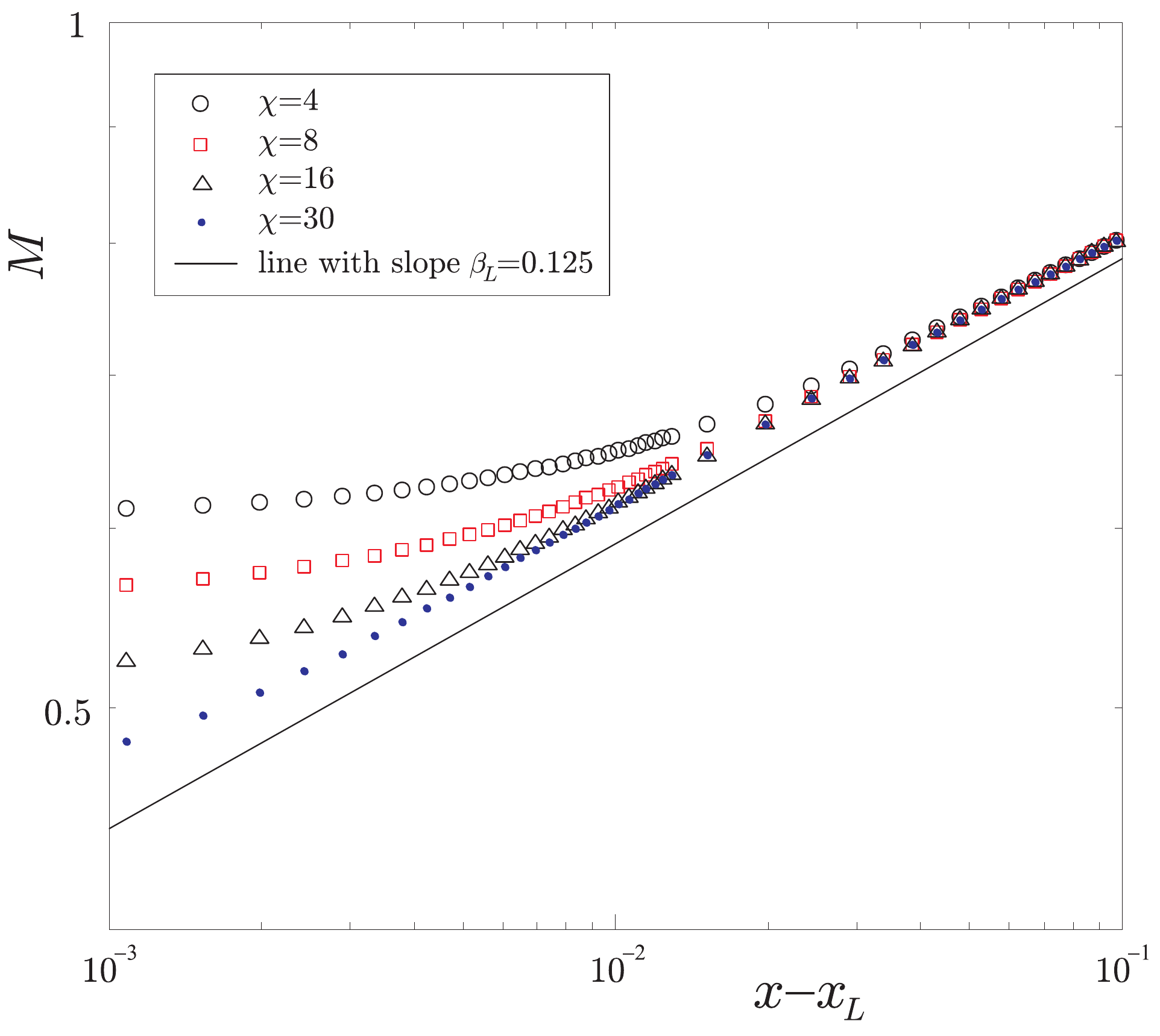}
	\caption{Transverse Ising model on an infinite line. Log-log plot of magnetization vs. $x-x_L$. 
	We also plot a line with slope $\beta_L = 0.125$.
	}
	\label{lineMexp}
	\end{center}
\end{figure}
In FIG.\ref{lineMexp}, we plot $M$ vs. $x-x_L$ on a log-log scale. We also plot a line with slope $0.125$. Observe that as $\chi$ increases, the data is better represented by a line down to smaller values of $x-x_L$. 
For the largest $\chi$ we display, a straight line fit of the data between $4\times 10^{-3} \leq x-x_L \leq 10^{-1}$ gives us a slope of $0.120$. Note that the mean field value of the critical exponent for magnetization is $0.5$.

\begin{figure}
	\begin{center}
	\includegraphics[width=5in]{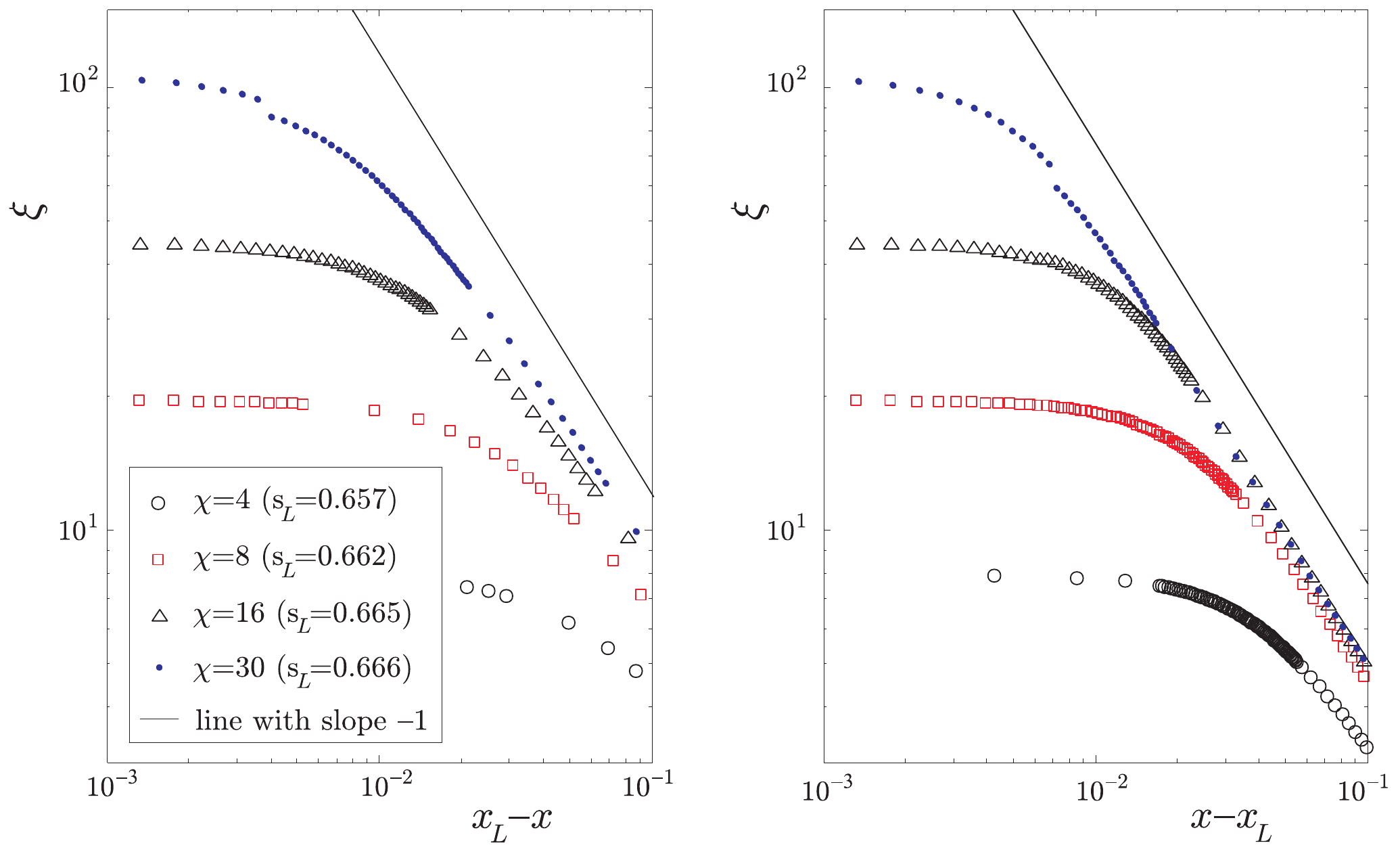}
	\caption{Transverse Ising model on an infinite line. Log-log plot of the correlation length vs. $|x-x_L|$. 
	We also plot a line with slope $-\nu_L = -1$.
	For each $\chi$ in the plot, we choose a numerical value of the critical point $s_L$ 
	as the point at which the correlation length is maximal.
	}
	\label{lineCexp}
	\end{center}
\end{figure}
The correlation function $\corrf$ 
can be computed efficiently using \eqref{twoexpect}.
Away from criticality, it falls off
exponentially as $e^{- |i-j|/\xi}$.
The falloff of the correlation function is necessarily exponential as long as $\mu_2$,
the second eigenvalue of $B$, is less than $1$.
The exact solution for the correlation length $\xi$ near the critical point has the form
\begin{eqnarray}
  \xi \propto |x-x_L|^{-1}
  \label{xidiverge}
\end{eqnarray}
as $x$ approaches $x_L$.
Already at low $\chi$ the iTEBD method captures the divergence of the correlation length. In FIG. \ref{lineCexp}, we plot $\xi$ vs. $|x-x_L|$ on a log-log plot, together with a line with slope $-1$.
Again, as $\chi$ increases, the data is better represented by a line closer to the phase transition.
For the highest $\chi$ we display, a straight line fit of the data between $2\times 10^{-2} \leq x_L-x \leq 4\times 10^{-1}$ gives us a slope of $-0.92$.
Note that the mean-field value of the critical exponent for the correlation length is $0.5$ (corresponding to slope $-0.5$ in the graph).


\subsection{The Infinite Tree (Bethe Lattice).}
\label{NRsection2}

The computational cost of the tree simulation is more expensive with growing $\chi$ than the line simulation, so we give our results only up to $\chi=8$. We run the imaginary time evolution with 10000 iterations (each iteration followed by several normalization steps) for each point $s$, taking a lower $\chi$ result as the starting point for the procedure. We also add a small symmetry-breaking longitudinal field with magnitude $B_z = 10^{-8}$. 

\begin{figure}
	\begin{center}
	\includegraphics[width=5.5in]{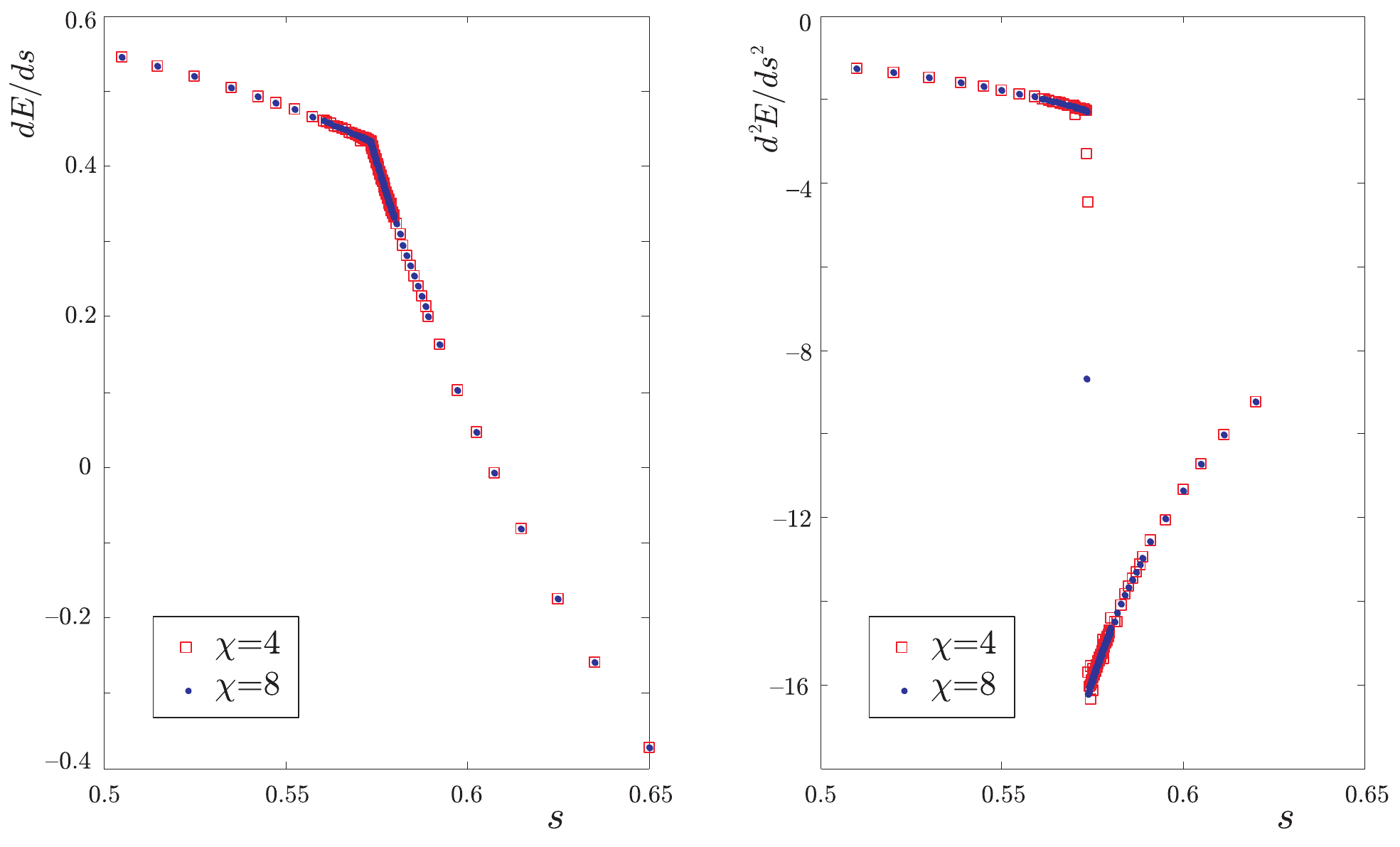}
	\caption{Transverse Ising model on an infinite tree. The first and second derivative with respect to $s$ of the ground state energy obtained via MPS.
	}
	\label{treeEE}
	\end{center}
\end{figure}

We see that the energy and its first derivative with respect to $s$ are continuous. 
However, we now observe a finite discontinuity in the second derivative of the
ground state energy (see FIG. \ref{treeEE}), as opposed
to the divergence on the infinite line.
This happens near $s=s_T \approx 0.5733$.

\begin{figure}
	\begin{center}
	\includegraphics[width=4.5in]{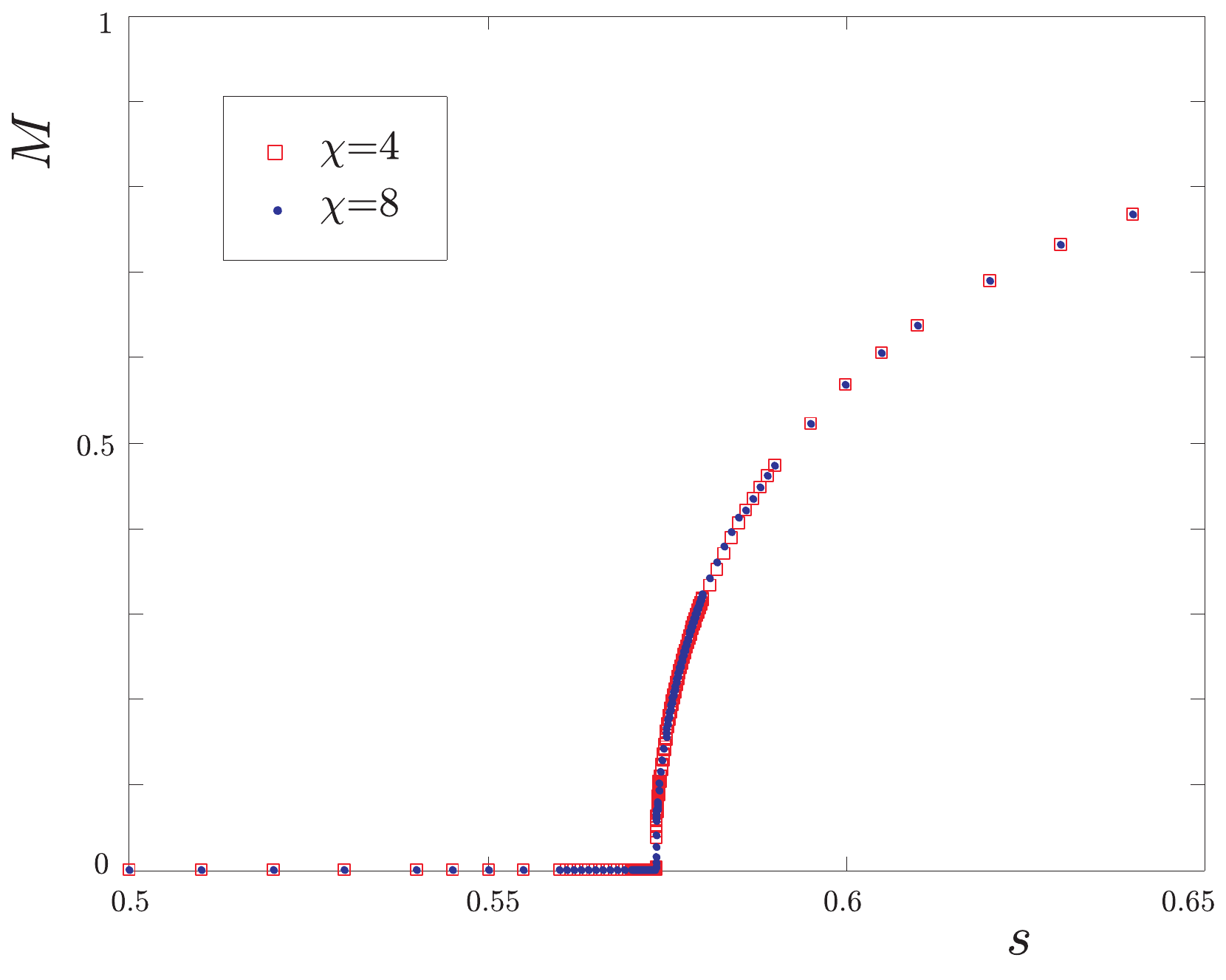}
	\caption{Transverse Ising model on an infinite tree. Magnetization vs. $s$, with $B_z=10^{-8}$.
	}
	\label{treeM}
	\end{center}
\end{figure}
\begin{figure}
	\begin{center}
	\includegraphics[width=4.5in]{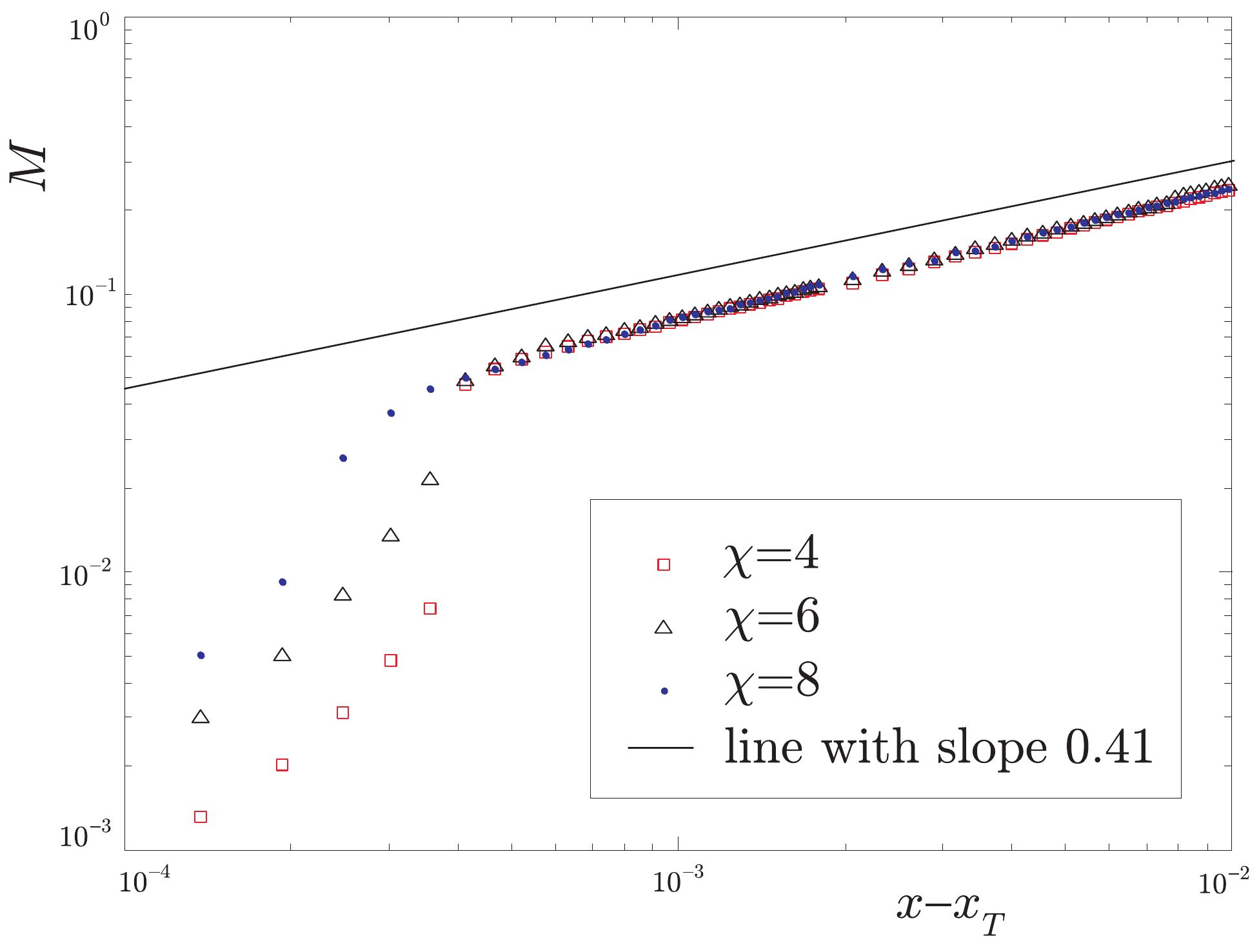}
	\caption{Transverse Ising model on an infinite tree. Log-log plot of magnetization vs. $x-x_T$, with $B_z = 10^{-8}$. 
	We also plot a line with slope $0.41$.
	}
	\label{treeMexp}
	\end{center}
\end{figure}

Similarly to the one-dimensional case, the magnetization quickly grows for $s>s_T$,
while it has (nearly) zero value for $s<s_T$ (see FIG.
\ref{treeM}). 
In FIG.\ref{treeMexp}, we plot the magnetization vs. $x-x_T$ on a log-log scale, where
$x$ is 
\begin{eqnarray}
	x = \frac{s}{3(1-s)},
\end{eqnarray}
with the value $x=x_T\approx 0.451$ at the phase transition (where $s=s_T\approx 0.5733$). 
We want to test whether the magnetization behaves like
\begin{eqnarray}
	M\propto (x-x_T)^{\beta}
\end{eqnarray}
for $x$ close to $x_T$, which would appear as a line on the log-log plot. 
As $\chi$ grows, the data is better represented by a straight line closer to the phase transition. 
If we fit the $\chi=8$ data for
$4\times 10^{-4} \leq x-x_L 4\times \leq 10^{-3}$, we get $\beta=0.41$. We add a line
with this slope to our plot.
Note that the mean-field value for the exponent $\beta$ is $0.5$, just as it is for the infinite line.

\begin{figure}
	\begin{center}
	\includegraphics[width=5in]{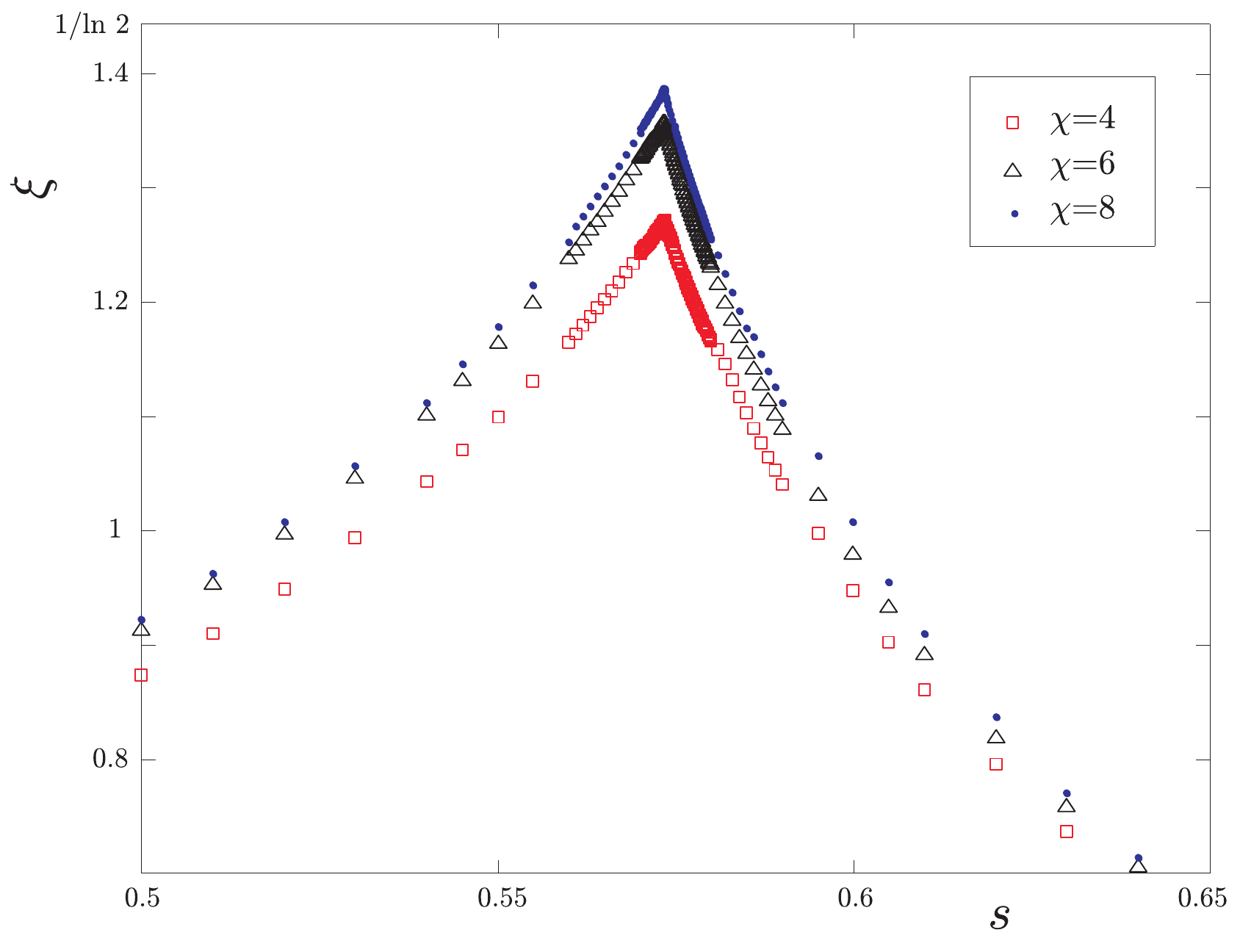}
	\caption{Transverse Ising model on an infinite tree. A linear plot of the correlation length vs. $s$.}
	\label{treeCexp}
	\end{center}
\end{figure}

We observe that the correlation length now rises up only to a finite value
(see FIG. \ref{treeCexp}).
As we increase $\chi$, the second eigenvalue of the $B$ matrix, $\mu_2$, approaches a maximum value close to $\frac{1}{2}$.
We conjecture that the limiting value of $\mu_2$ is indeed $\frac{1}{2}$, which corresponds to a finite correlation length with value $(\ln 2)^{-1}$. Note that for the infinite line, the second eigenvalue of $B$ approaches $1$, and so the correlation length is seen to diverge at the phase transition.


\section{The Not $00$ Model}
\label{not00section}

We now look at a model with a different interaction term.
Starting with an antiferromagnetic interaction,
we add a specific longitudinal field at each site. As in the previous section, we parametrize our Hamiltonian \eqref{ourH00} with a single parameter $s$:
\begin{eqnarray}
	H_{\textsc{not}\,00} &=& s \sum_{\langle i,j\rangle} 
		\underbrace{\frac{1}{4}\left( 1 + \sigma_z^i + \sigma_z^j + \sigma_z^i \sigma_z^j  \right)}_{H_{ij}}
		+ \frac{b(1-s)}{2} \sum_i \left(1-\sigma_x^{i}\right), \label{Hnot00}
\end{eqnarray}
with $b=2$ on the line and $b=3$ on the tree.
We choose the longitudinal field in such a way that the nearest-neighbor interaction term $H_{ij}$ becomes a projector, expressed in the computational basis as  
\begin{eqnarray}
	H_{ij} = \ket{00}\bra{00}_{ij},
\end{eqnarray}
thus penalizing only the $\ket{00}$ configuration of neighboring spins. Accordingly, we call this model \textsc{not} 00. The ground state of the transverse Ising model \eqref{ourH} at $s=1$ has degeneracy 2. For \eqref{Hnot00} on the infinite line or the Bethe lattice, the degeneracy of the ground state at $s=1$ is infinite, as any state that does not have two neighboring spins in state $\ket{0}$ has zero energy.

\subsection{Infinite Line}
We use our numerics to investigate the properties of \eqref{Hnot00} 
on the infinite line as a function of $s$.
\begin{figure}
	\begin{center}
	\includegraphics[width=6in]{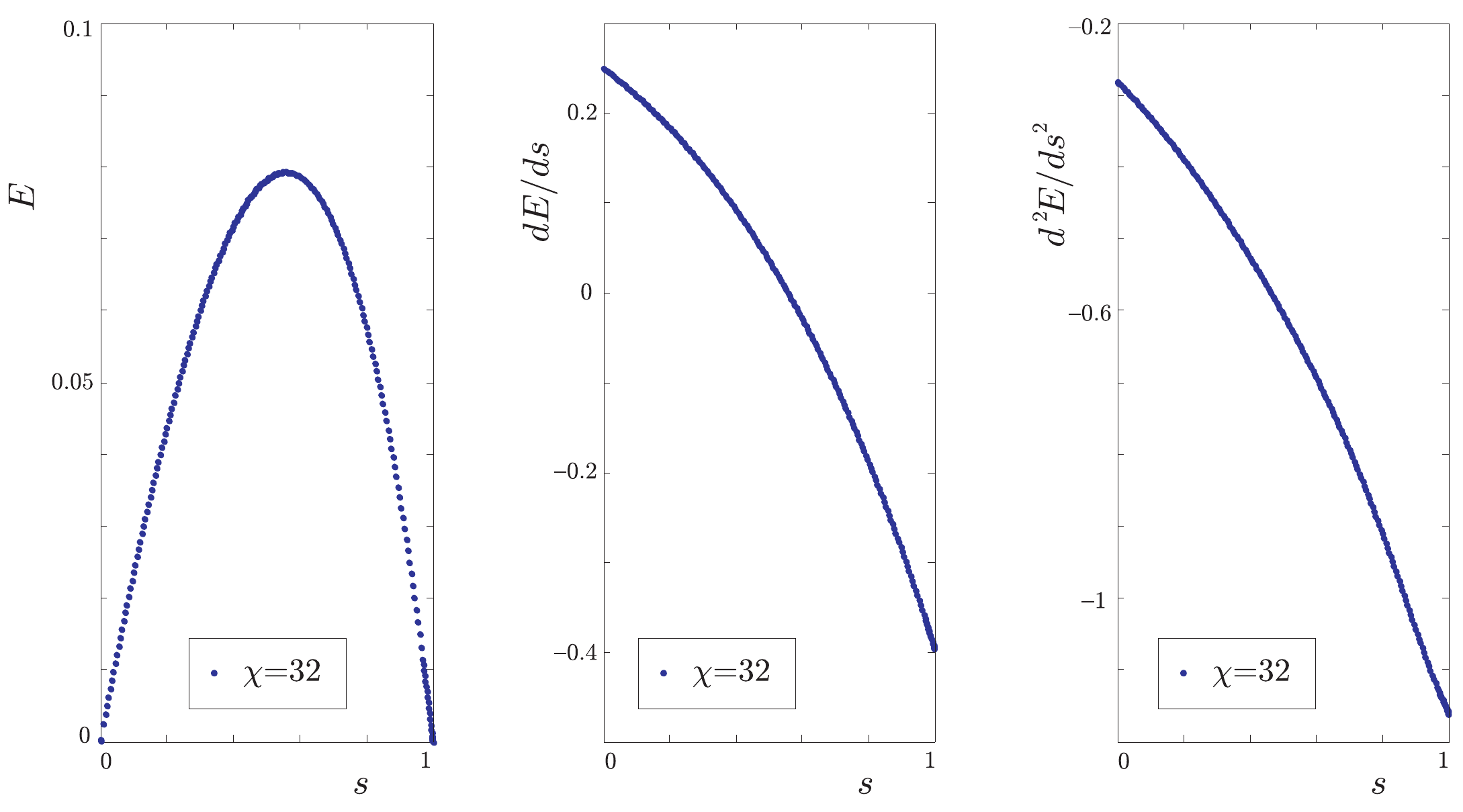}
	\caption{The \textsc{not} $00$ model on an infinite line. The ground state energy and its first two derivatives with respect to $s$.}
	\label{fig00lineE3}
	\end{center}
\end{figure}
\begin{figure}
	\begin{center}
	\includegraphics[width=5in]{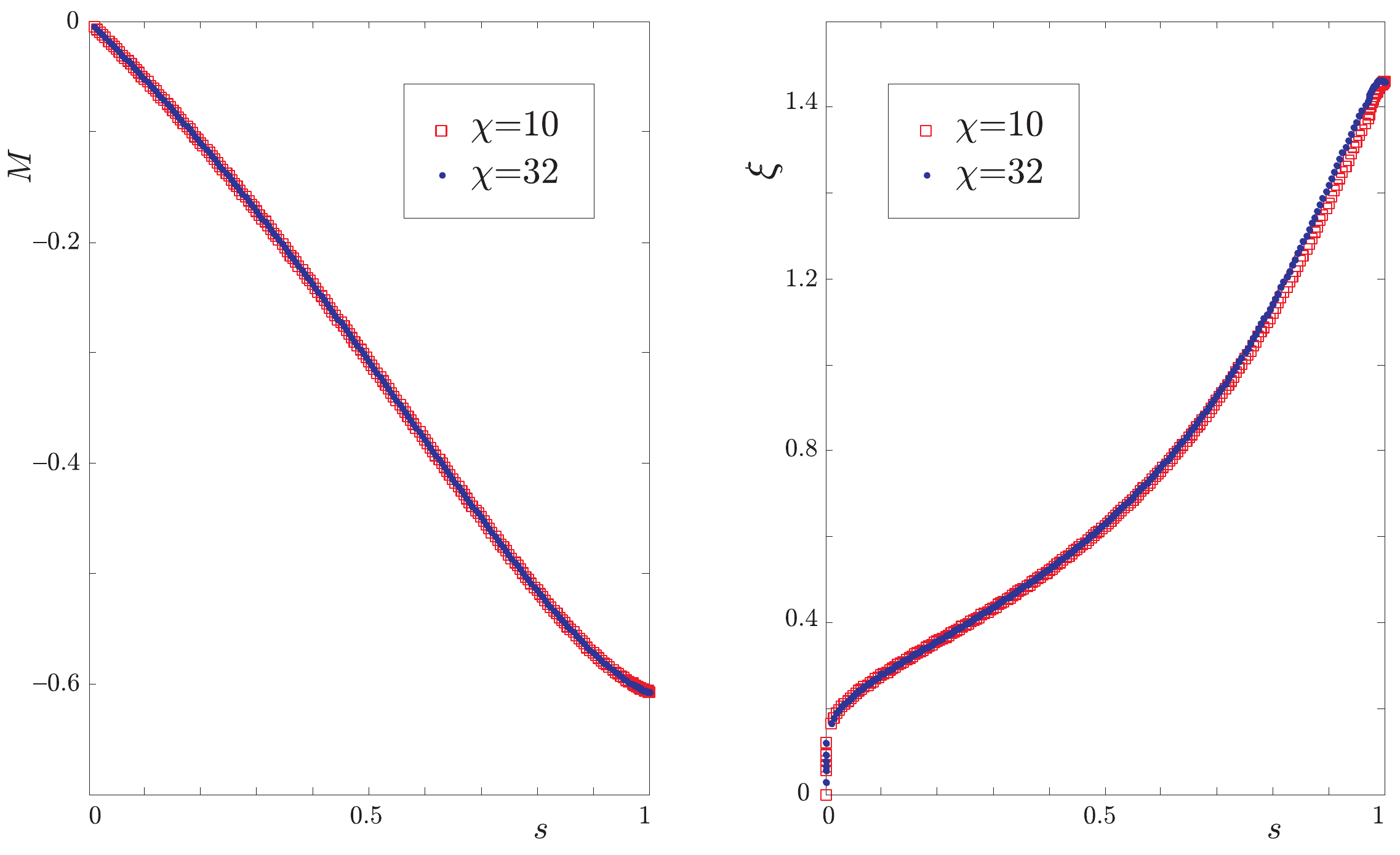}
	\caption{The \textsc{not} $00$ model on an infinite line. Magnetization as a function of $s$ and correlation length as a function of $s$.}
	\label{fig00lineM}
	\end{center}
\end{figure}
Our numerical results show continuous first and second derivatives of the energy 
with respect to $s$ (see FIG.\ref{fig00lineE3}). The magnetization $M=\means{\sigma_z}$ decreases 
continuously and monotonically from $0$ at $s=0$ to a final value of $-0.606$ at $s=1$ (see FIG.\ref{fig00lineM}). 
The second eigenvalue of the $B$ matrix \eqref{secondeig} rises continuously from $0$ at $s=0$,
approaching $0.603$ at $s=1$ (see FIG.\ref{fig00lineM}). Because $\mu_2 < 1$, the 
correlation length $\xi$ is finite for all values of $s$ in this case.
These results imply that there is no phase transition for this model as we vary $s$.

\begin{figure}
	\begin{center}
	\includegraphics[width=4in]{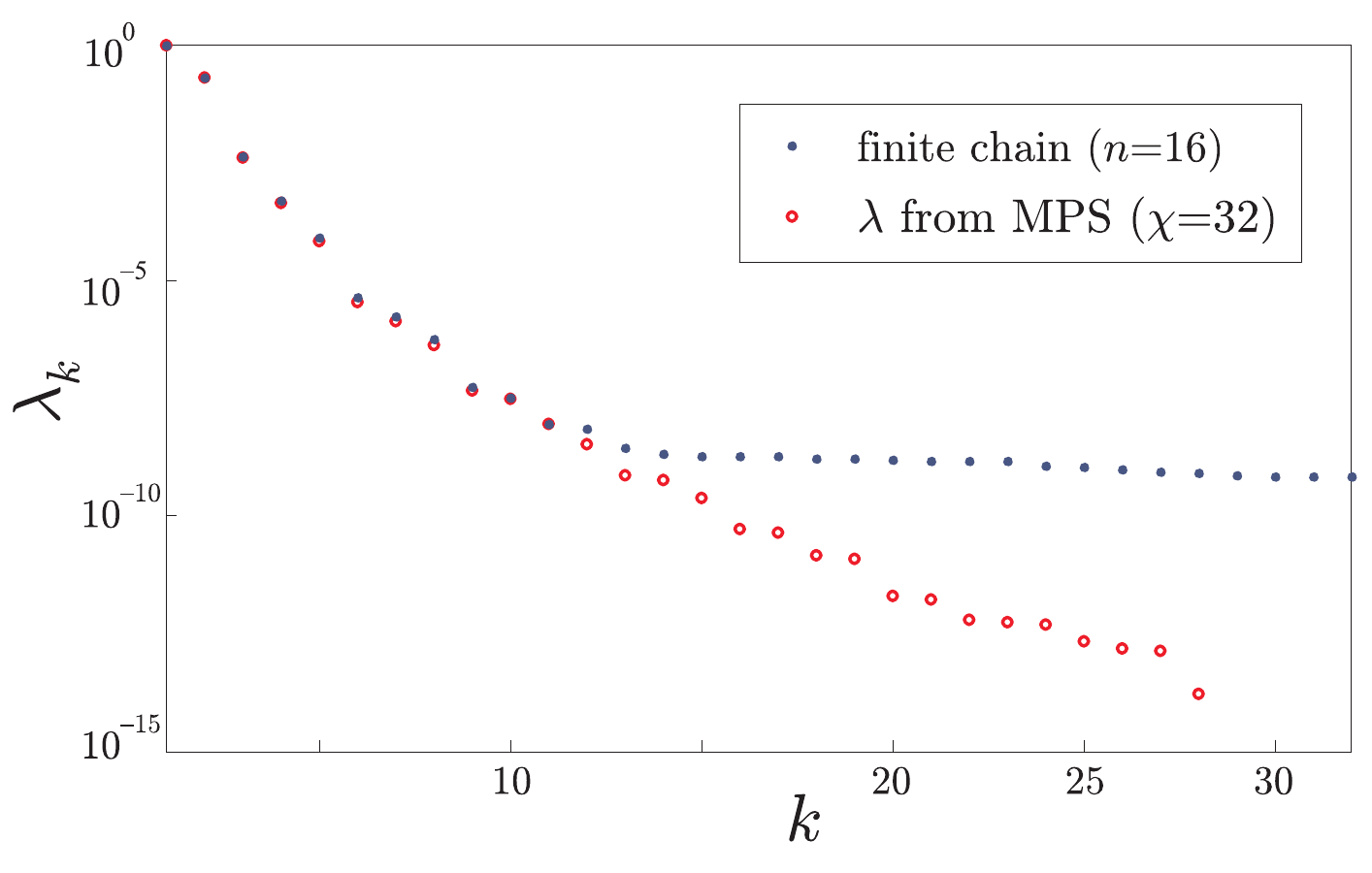}
	\caption{Ground state of the \textsc{not} $00$ model on a line at $s=1$. Comparison of the 	exact Schmidt coefficients for a division across the middle of a finite chain
	 and of the MPS values ($\chi=32$) for an infinite line.}
	\label{not00lamfig}
	\end{center}
\end{figure}

As a test of our results, we compute the magnetization at $s=1$ exactly for this model on a finite chain (and ring) of up to $n=17$ spins. We maximize the expectation value of $H_B = \sum_i \sigma_x^i$ within the subspace of all allowed states at $s=1$ (with no two zeros on neighboring spins), thus minimizing the expectation value of the second term in \eqref{Hnot00} for $s$ approaching 1. 
We compute the magnetization $M = \langle \sigma_z^{i}\rangle$
for the middle $i = \lfloor \frac{n}{2} \rfloor$ spin for the ground state of the \textsc{not} 00 model 
exactly
for a finite chain and ring of up to $17$ spins at $s=1$. As we increase $n$, the value of $M$ converges to $-0.603$ (much faster for the ring, as the values of $M$ for $n=14,17$ differ by less than $10^{-4}$). Recall that we obtained $M=-0.606$ from our MPS numerics with $\chi = 16$ for the \textsc{not} 00 model on an infinite line.
We also compare the values of the Schmidt coefficients across the central division of the finite chain ($n=16$) to the elements of the $\lambda$ vector obtained using our MPS numerics with $\chi=32$. We observe very good agreement for the 11 largest values of $\lambda_k$, with the difference that our MPS values keep decreasing (exponentially), while the finite-chain values flatten out at around $\lambda_{k>14} \approx 10^{-9}$ (see FIG.\ref{not00lamfig}). The behavior of the components of $\lambda$ from MPS doesn't change with increasing $\chi$.

In the ground state of \eqref{Hnot00} at $s=1$, the overlap with the $\ket{00}$ state of any two neighboring spins is exactly $0$. If the $\Gamma$ tensors are the same at every site, the component of the state $\ket{\psi}$ that has overlap with the state $\ket{00}$ on nearest neighbors can be expressed as 
\begin{eqnarray}
	\sum_{a,b,c}   
	\left( \lambda_a \Gamma_{ab}^{0} \lambda_b \Gamma_{bc}^{0} \lambda_c \right) 
	\ket{\phi_a}\ket{00}\ket{\phi_c}.
\end{eqnarray}
Furthermore, when the $\Gamma$ tensors are symmetric, the elements of the $\lambda$ vectors must be allowed to take negative values
to make this expression equal to zero. Note that until now, we used only positive $\lambda$ vectors, knowing that they come from Schmidt decompositions, which give us the freedom to choose the components of $\lambda$ to be positive and decreasing. 

The negative signs in the $\lambda$ vector can be absorbed into every other $\Gamma$ tensor, resulting in a state with two different $\Gamma$ (for the even and odd-numbered sites) and only positive $\lambda$'s. In fact, this is what we observe in our numerics, which assume positive $\lambda$, but allow two different $\Gamma$ tensors (see \ref{TIline}). If we allow the elements of $\lambda$ to take negative values, our numerically obtained $\Gamma$ tensors are identical.

\subsection{Infinite Tree}
Here, we numerically investigate the \textsc{not} $00$ model \eqref{Hnot00} on the Bethe lattice. 
\begin{figure}
	\begin{center}
	\includegraphics[width=6in]{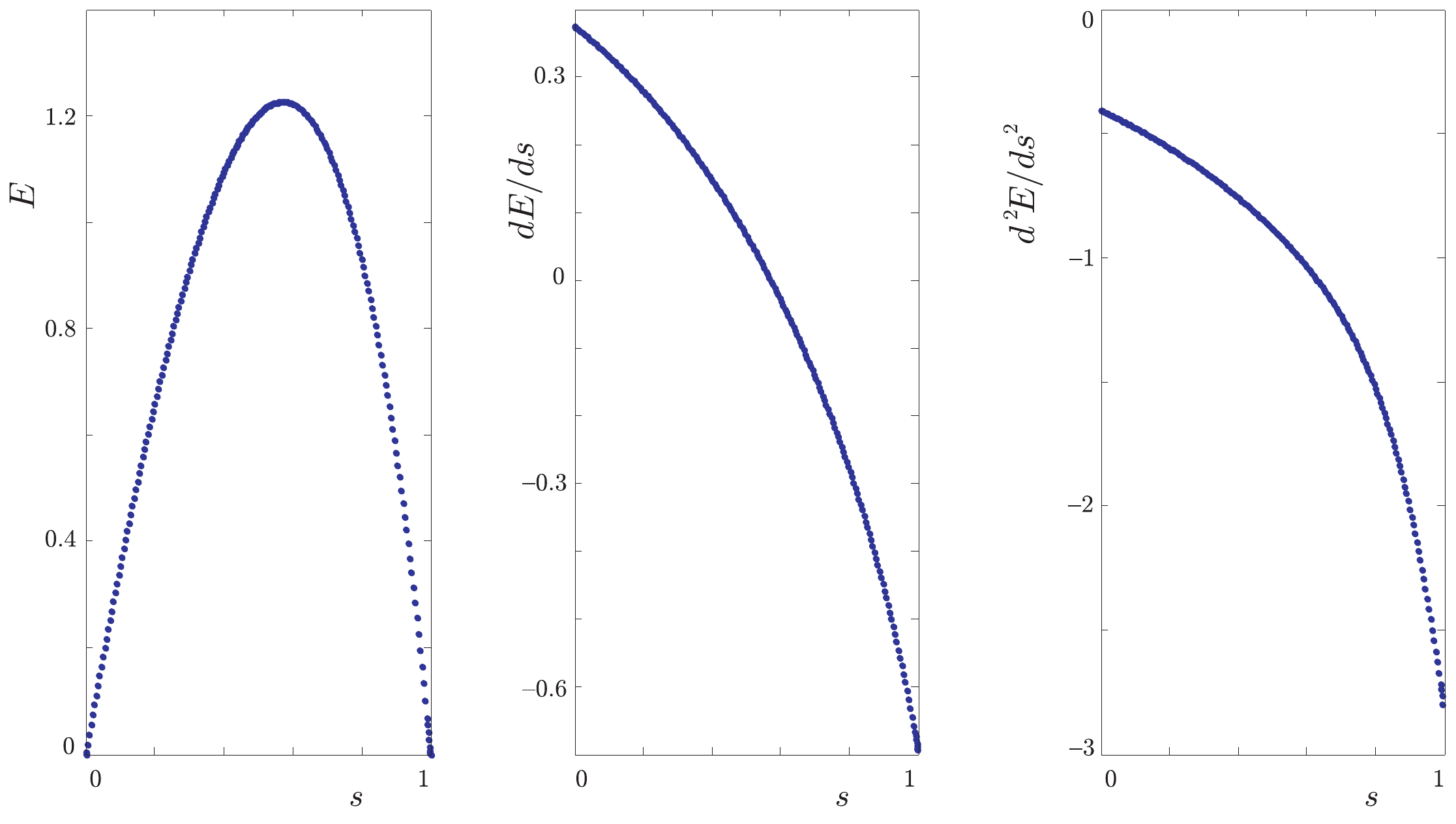}
	\caption{The \textsc{not} $00$ model on an infinite tree. The ground state energy and its first two derivatives with respect to $s$.}
	\label{fig00treeE3}
	\end{center}
\end{figure}
\begin{figure}
	\begin{center}
	\includegraphics[width=5in]{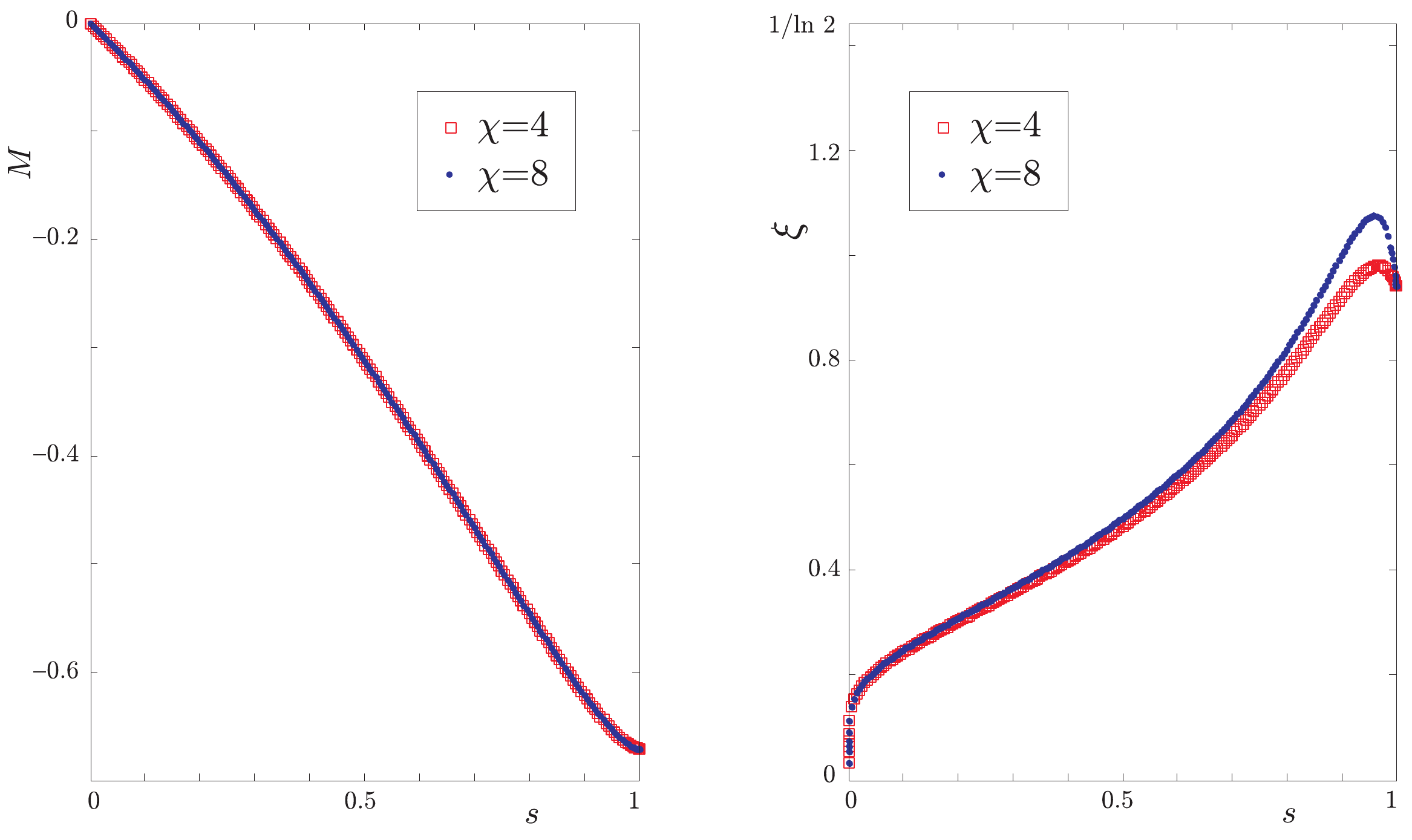}
	\caption{The \textsc{not} $00$ model on an infinite tree. Magnetization as a function of $s$ and correlation length as a function of $s$.}
	\label{fig00treeM}
	\end{center}
\end{figure}
As on the line, the numerics show continuous first and second derivatives of the energy with respect to $s$ (see FIG.\ref{fig00treeE3}) and a continuous decrease in the magnetization from $0$ at $s=0$ to $-0.671$ at $s=1$ (see FIG.\ref{fig00treeM}). The correlation length behaves similarly as on the line, increasing with $s$, but it reaches a maximum at $s=0.96$ for $\chi=8$. The maximum value of $\xi$ is apparently lower than $1/\textrm{ln}\,2$, (see FIG.\ref{fig00treeM}), meaning that 
on the tree, the correlation function $\corrf$ falls off with distance faster than $2^{-|i-j|}$ for all $s$. 



\section{Stability and Correlation Lengths on the Bethe Lattice}
\label{stabilitysection}

We have found that on the Bethe lattice, for both our models the second eigenvalue 
$\mu_2$ of the matrix $B$ \eqref{secondeig}, which determines the correlation length, apparently is never greater than $\frac{1}{2}$. In this section we argue that this is a model-independent, and calculation method independent, consequence of assuming that a translation-invariant ground state is the stable limit of a sequence of ground states of finite Cayley trees as the size of the tree grows. 
For a related problem, the stability of recursions for Valence Bond States on Cayley trees has been investigated by Fannes et.al. in \cite{VBScayley}.

\begin{figure}
	\begin{center}
	\includegraphics[width=2in]{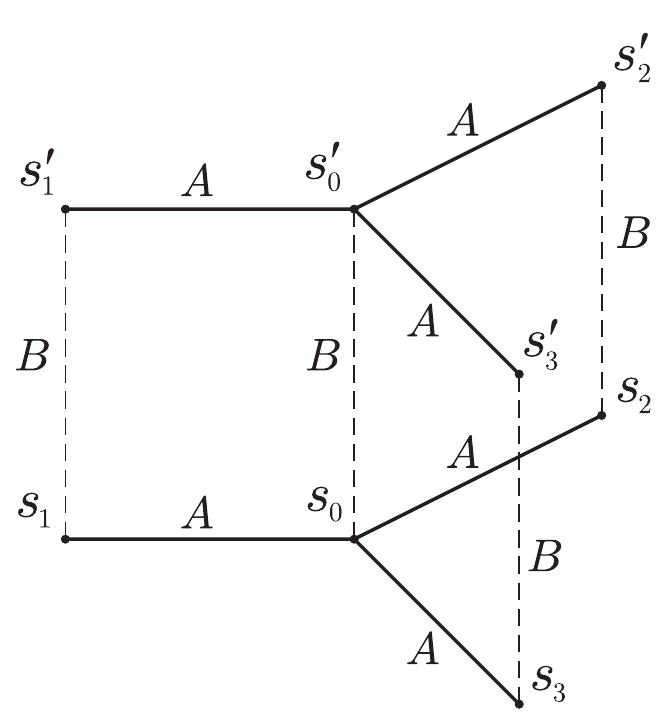}
	\caption{A system of classical spins on a lattice whose layers are Cayley trees. $A$ and $B$ denote the Boltzmann factors.}
	\label{figboltzmann}
	\end{center}
\end{figure}

The Hamiltonians \eqref{ourH1} and \eqref{ourH00} each consist of sums of terms 
$H^{(x)}_k$, $H^{(z)}_m$, as in \eqref{trotter}, where each term
$H^{(x)}_k$ depends on a single $\sigma_x$ and each $H^{(z)}_m$ on a neighboring pair of $\sigma_z$. We calculate the quantum partition function
\begin{eqnarray}
	Z(\beta) = \tr e^{-\beta H}
	\label{partf1}
\end{eqnarray}
as the limit of 
\begin{eqnarray}
	Z(N,\Delta t) = \tr \left[ 
		\prod_k e^{-\Delta t H_k^{(x)} }
		\prod_m e^{-\Delta t H_m^{(z)}} \right]^N
	\label{partf2}
\end{eqnarray}
as $\Delta t \rightarrow 0$, $N \rightarrow \infty$ with $N\Delta t = \beta$. To find the properties of the ground state, we take $\beta \rightarrow \infty$ so that we need $Z(N,\Delta t)$ as $\Delta t \rightarrow 0$, $N \rightarrow \infty$ with $N\Delta t \rightarrow \infty$ and $N(\Delta t)^3 \rightarrow 0$ (to make the error in using the Trotter-Suzuki formula go to zero). We interpret \eqref{partf2} as giving the {\em classical} partition function of a system of Ising spins ($s=\pm 1$) on a lattice consisting of $N$ horizontal layers, each of which is a Cayley tree of radius $M$ (i.e with a central node and concentric rings of $3, 3\times2, 3\times 2^2, \dots, 3\times 2^{M-1}$ nodes). 
We can write \eqref{partf2} as
\begin{eqnarray}
	Z(N,\Delta t) = \sum_{\{s\}} \prod A \prod B,
	\label{Zprod}
\end{eqnarray}
where the sum is over all configurations of $N\times(3\times 2^M-2)$ spins $s=\pm 1$
and the products are of a Boltzmann factor $A$ for each horizontal link in the Cayley trees, and a Boltzmann factor $B$ for each vertical link between corresponding nodes in neighboring layers
(see FIG.\ref{figboltzmann})
(layer $N$ is linked to layer $1$ to give the trace).
The factor $A$ for the link between nodes $i,j$ in the same horizontal layer is given by  
\begin{eqnarray}
	A(s_i,s_j) &=& e^{-\Delta t H^{(z)}_{(ij)}(s_i,s_j)}.
	\label{defA} 
\end{eqnarray}
The factor $B$ for the link between nodes $i,i'$ in the same vertical column is given by
\begin{eqnarray}
	B(s_i,s_i') &=& \bra{\sigma_z=s_i'} e^{-\Delta t H^{(x)}_{(i)}} \ket{\sigma_z=s_i}.
	\label{defB}
\end{eqnarray}
When all the terms $H^{(z)}_{(ij)}$ are of the same form, as are all the terms $H^{(x)}_{(i)}$, the form of the factors $A$ and $B$ does not depend on which particular links they belong to.

Each term in the sum, divided by $Z$, can be thought of as the probability of a configuration $\{s\}$. In what follows we will keep $N$ and $\Delta t$ fixed and consider the limit $\Mlim$, i.e. finite Cayley tree $\rightarrow$ Bethe lattice. We will then suppose that our results, which are independent of the form of $A$ and $B$ (provided $A,B > 0$) will also hold after the $N\rightarrow \infty$ limit is taken, i.e. for the quantum ground state.

\begin{figure}
	\begin{center}
	\includegraphics[width=3in]{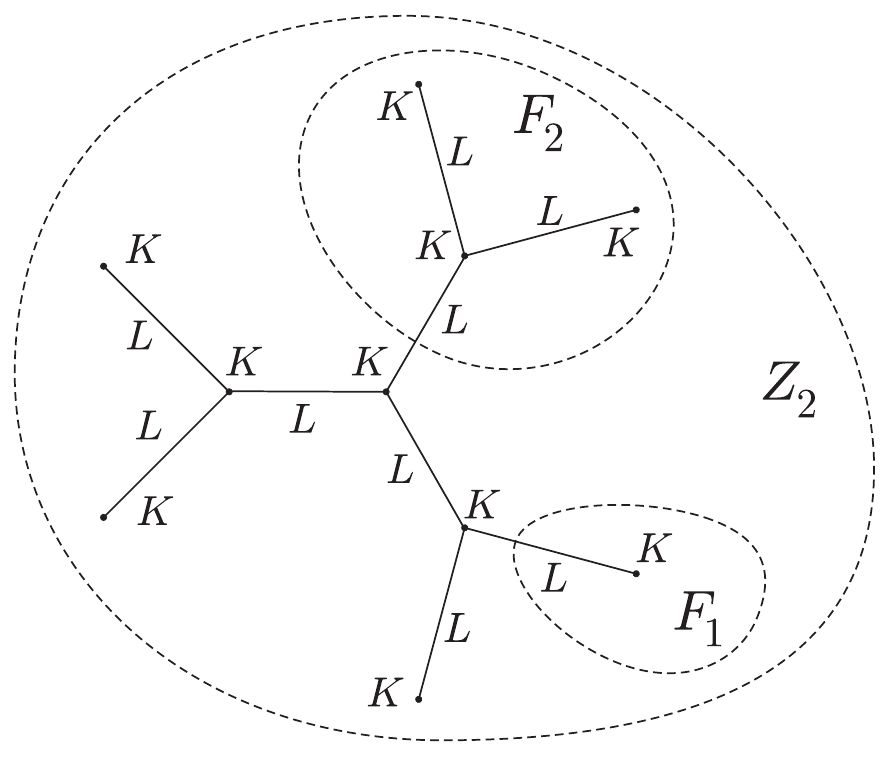}
	\caption{The $M=2$ Cayley tree.}
	\label{figm2tree}
	\end{center}
\end{figure}

We will think of our lattice as a single tree, with each node being a vertical column of $N$ spins. (A recent use of this technique to investigate the quantum spin glass on the Bethe lattice is in \cite{LSSspinglass}.)
Denote by $\vec{s}$ the vector of $N$ values of $s$ along a column. Let $K(\vec{s})$ be the
product of the $N$ factors $B(s,s')$ along a column and let $L(\sve,\svp)$ be the product of 
$N$ factors $A(s,s')$ on the horizontal links between nearest neighbor columns. Let $Z_M$
be the partition function for a tree of radius $M$. 
We can calculate $Z_M$ by a recursion on $M$ as follows:
\begin{eqnarray}
	Z_M &=& \sum_{\sve} K(\sve) [ F_M(\sve) ]^3, \label{Zdefinition} \\
	F_M(\sve) &=& \sum_{\svp} L(\sve,\svp) K(\svp) [ F_{M-1}(\svp) ]^2, 	
			\label{Frecursion} \\
	F_0(\sve) &=& 1.	
			\label{Finit} 
\end{eqnarray}
It is easy to see that this recursion gives the correct $Z_M$ (the case $M=2$ is shown in
 FIG.\ref{figm2tree}).
We also see that
\begin{eqnarray}
	P_M(\sve) &=& \frac{1}{Z_M} K(\sve) [ F_{M}(\sve) ]^3 
			\label{probability} 
\end{eqnarray}
is the probability of the configuration $\sve$ along the central column.
For the case $N=1$, i.e. classical statistical mechanics on a tree, this is the well-known method to find an exact solution \cite{classicalexact}.

In order to have a well-defined translationally invariant limit as $\Mlim$ we would like the
recursion \eqref{Frecursion} for $F_M$ to have an attractive fixed point $F$ which $F_M$
approaches as $\Mlim$.
`Attractive' means that if we start the recursion with a different $F_0(\sve)$, sufficiently close to $F_0(\sve)=1$, the limiting value of $F_M(\sve)$ will be the same fixed point. This in turn implies that on a Cayley tree with large $M$, small changes in the Hamiltonian on the outer edge will have small effects on the properties of the central region.

First however we need to fix the overall normalization of $F_M(\sve)$, since if $F_M(\sve)$ satisfies
\eqref{Frecursion}, so does $a^{2^M} F_M(\sve)$ which rules out an attractive fixed point.

Let
\begin{eqnarray}
	F_M(\sve) &=& Z_M^{\frac{1}{3}} \hat{F}_{M}(\sve),
			\label{Fhat} 
\end{eqnarray}
so that
\begin{eqnarray}
	\sum_{\sve} K(\sve) [ \hat{F}_{M}(\sve) ]^3 &=& 1,
			\label{Fhatnorm} 
\end{eqnarray}
and
\begin{eqnarray}
	P_M(\sve) &=& K(\sve) [ \hat{F}_{M}(\sve) ]^3.
			\label{Phat} 
\end{eqnarray}
The recursion relation becomes
\begin{eqnarray}
	\hat{F}_M(\sve) &=& \lambda_M \sum_{\svp} L(\sve,\svp) K(\svp) 
		[ \hat{F}_{M-1}(\svp) ]^2, 	
			\label{Fhatrecursion}
\end{eqnarray}
with $\lambda_M$ determined by the normalization condition \eqref{Fhatnorm}. 
We can now suppose that
\begin{eqnarray}
	\hat{F}_M(\sve) \rightarrow \hat{F}(\sve) \quad \textrm{as} \quad \Mlim, 	
			\label{Fgoes} 
\end{eqnarray}
with
\begin{eqnarray}
	\hat{F}(\sve) &=& \lambda \sum_{\svp} L(\sve,\svp) K(\svp) [ \hat{F}(\svp) ]^2, 	
			\label{Flimit} 
\end{eqnarray}
and
\begin{eqnarray}
	\sum_{\sve} K(\sve) [ \hat{F}(\sve) ]^3 &=& 1.
			\label{Flimnorm} 
\end{eqnarray}
To determine whether $\hat{F}$ is an {\em attractive} fixed point, let
\begin{eqnarray}
	\hat{F}_M(\sve) &=& \hat{F}(\sve) + f_M(\sve), \label{Fperturb} \\
	\lambda_M &=& \lambda (1 + \ep_M), \label{lperturb}
\end{eqnarray}
with $f_M \rightarrow 0$ and $\ep_M \rightarrow 0$ as $\Mlim$.
To first order in $f_M$, $\ep_M$, \eqref{Fhatrecursion} and \eqref{Fhatnorm} become
\begin{eqnarray}
	f_M(\sve) &=& \ep_M \hat{F}(\sve) + 2 \sum_{\svp} T(\sve,\svp) f_{M-1}(\svp),
		\label{frecursion} \\
	\sum_{\sve} K(\sve) [ \hat{F}(\sve)]^2 f_M(\sve) &=& 0,
		\label{frecursion2}	
\end{eqnarray}
where
\begin{eqnarray}
	T(\sve,\svp) = \lambda L(\sve,\svp) K(\svp) \hat{F}(\svp).
			\label{Tdef} 
\end{eqnarray}
From \eqref{Flimit},
\begin{eqnarray}
	\sum_{\svp}  T(\sve,\svp) \hat{F}(\svp) = \hat{F}(\sve),
	\label{TFhat}
\end{eqnarray}
and since $L(\sve,\svp) = L(\svp,\sve)$,
\begin{eqnarray}
	\sum_{\svp}  K(\svp) [ \hat{F}(\svp) ]^2 T(\svp,\sve) 
		= K(\sve) [\hat{F}(\sve) ]^2,
\end{eqnarray}
i.e. the linear operator $T$ has an eigenvalue one, with right eigenvector $\hat{F}$ and left eigenvector $K\hat{F}^2$ (which from \eqref{Flimnorm} have scalar product one).
\eqref{frecursion} now gives
\begin{eqnarray}
	\sum_{\sve} K(\sve) [ \hat{F}(\sve) ]^2 f_M(\sve) 
		= \ep_M  + 2\sum_{\sve} K(\sve) [ \hat{F}(\sve) ]^2 f_{M-1}(\sve),
\end{eqnarray}
so from \eqref{frecursion2}, $\ep_M = 0$. Let 
\begin{eqnarray}
	T^{\perp}(\sve,\svp) = T(\sve,\svp) - \hat{F}(\sve) K(\svp) [\hat{F}(\svp)]^2,
	\label{Tperp1}
\end{eqnarray}
so that 
\begin{eqnarray}
	\sum_{\svp} T^{\perp}(\sve,\svp) \hat{F}(\svp) &=& 0,
	\label{Tperp2}
\end{eqnarray}
and
\begin{eqnarray}
	\sum_{\svp} K(\svp) [ \hat{F}(\svp) ]^2 T^{\perp}(\svp,\sve) 
		= 0.
	\label{Tperp2b}
\end{eqnarray}
\eqref{frecursion} now becomes 
\begin{eqnarray}
	f_M(\sve) = 2 \sum_{\svp} T^{\perp}(\sve,\svp) f_{M-1}(\svp).
		\label{ffromT}
\end{eqnarray}
\eqref{ffromT} shows that $\hat{F}(\sve)$ is an attractive fixed point if and only if 
\begin{eqnarray}
	\norms{T^{\perp}} < \frac{1}{2}
	\label{Tbound}
\end{eqnarray}
(for a tree with valence $p+1$ at each vertex, $\frac{1}{2}$ is replaced by $\frac{1}{p}$).

We can in fact prove that there does exist an $\hat{F}(\sve)$ satisfying \eqref{Flimit} and \eqref{Flimnorm}, for which the corresponding $T^\perp$ has a maximum eigenvalue less than $\frac{1}{2}$. Define a function $\Phi$ of $\hat{F}(\sve)$ by
\begin{eqnarray}
	\Phi[\hat{F}] = \sum_{\sve,\svp} [\hat{F}(\sve)]^2 K(\sve)
		L(\sve,\svp)K(\svp)[\hat{F}(\svp)]^2.
\end{eqnarray}
We look for a maximum of $\Phi$ with $\hat{F}(\sve)$ restricted to the region
\begin{eqnarray}
	\sum_{\sve} K(\sve)[\hat{F}(\sve)]^3 &=& 1,\\
	\hat{F}(\sve) &\geq& 0.
\end{eqnarray}
A maximum must exist, but it might be on the boundary of the region, i.e. it might have $\hat{F}(\sve)=0$ for some values of $\sve$. Elementary calculations (omitted here) establish that stationary values of $\Phi$ on the boundary cannot be maxima. At stationary points 
in the interior of the region, i.e. with $\hat{F}(\sve)>0$ for all $\sve$, \eqref{Flimit} and \eqref{Flimnorm} must be satisfied. If such a stationary point is a maximum, all the eigenvalues of $1-2T^\perp$ are $\geq 0$, i.e. all the eigenvalues of $T^\perp$ are $\leq \frac{1}{2}$. This is weaker than the attractive fixed point condition, which also requires that no eigenvalue is less than $-\frac{1}{2}$, but does correspond to our observed property of $\mu_2$.

We now examine the joint probability distribution of $\vec{s}_0$ and $\vec{s}_d$, where $0$ denotes the central column and $d$ a column distance $d$ from the center.
\begin{figure}
	\begin{center}
	\includegraphics[width=4in]{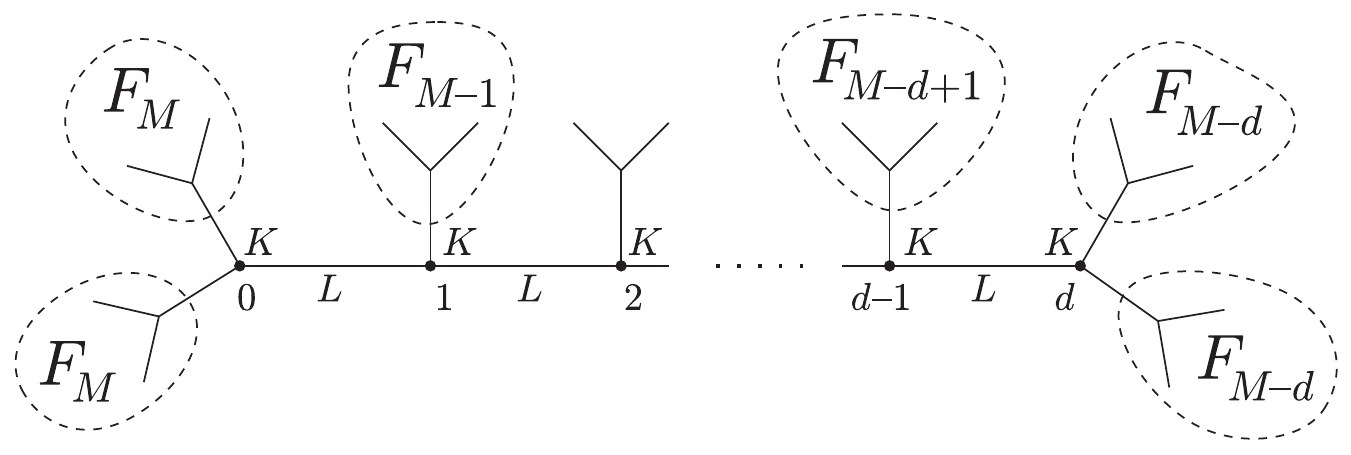}
	\caption{Computing the probability distribution of $\vec{s}_0$ and $\vec{s}_d$.}
	\label{figdistanced}
	\end{center}
\end{figure}
From FIG.\ref{figdistanced} we see that
\begin{eqnarray}
	P_M(\vec{s}_0,\vec{s}_d) = \frac{1}{Z_M} \sum_{\vec{s}_1,\cdots,\vec{s}_{d-1}}
		[ F_M(\vec{s}_0) ]^2 
			K(\vec{s}_0) L(\vec{s}_0,\vec{s}_1) 
			F_{M-1}(\vec{s}_1)
			\dots && \\
			\dots L(\vec{s}_{d-1},\vec{s}_d) K(\vec{s}_d)
			[ F_{M-d}(\vec{s}_d) ]^2.&& \nonumber
\end{eqnarray}
As $\Mlim$ (with $d$ fixed) this becomes (using \eqref{Tdef})
\begin{eqnarray}
	P(\vec{s}_0,\vec{s}_d) &=& 
		C [ \hat{F}(\vec{s}_0) ]^2 K(\vec{s}_0) 
		T^{d} (\vec{s}_0,\vec{s}_d)
		 \hat{F}(\vec{s}_d),
	 \label{PfromT}
\end{eqnarray}
where the normalization $C$ is determined by 
\begin{eqnarray}
	\sum_{\vec{s}_0,\vec{s}_d} P(\vec{s}_0,\vec{s}_d) = 1.
\end{eqnarray}
Using \eqref{TFhat} we find
\begin{eqnarray}
	P(\vec{s}_0) = \sum_{\vec{s}_d} P(\vec{s}_0,\vec{s}_d) 
		= C K(\vec{s}_0) [ \hat{F}(\vec{s}_0) ]^3,
\end{eqnarray}
so from \eqref{Flimnorm}, $C=1$ (and $P(\vec{s}_0)$ agrees with the limit of \eqref{Phat}). Expressing \eqref{PfromT} in terms of $T^{\perp}$ using \eqref{Tperp1}, \eqref{Tperp2} and \eqref{Tperp2b}, 
\begin{eqnarray}
	P(\vec{s}_0,\vec{s}_d) - P(\vec{s}_0) P(\vec{s}_d)
	 	&=& K(\vec{s}_0) [ \hat{F}(\vec{s}_0) ]^2
			(T^{\perp})^d (\vec{s}_0,\vec{s}_d)  \hat{F}(\vec{s}_d).
\end{eqnarray}
Thus the correlation between $\vec{s}_0$ and $\vec{s}_d$ falls off as $\mu^{d}$, where $\mu$ is the eigenvalue of $T^\perp$ with maximum modulus, and so from \eqref{Tbound}, faster than $1/2^d$.

If this conclusion is correct (and clearly the argument is less than rigorous), it establishes more than our experimental observation that $\mu_2<\frac{1}{2}$. The quantum limit of $P(\vec{s}_0,\vec{s}_d)$ encodes not only the static correlation 
$\bra{\psi_0} s_0 s_d \ket{\psi_0}$ in the ground state, but also the 
imaginary time dependent correlation
$\bra{\psi_0}e^{Ht}s_0 e^{-Ht} s_d \ket{\psi_0}$ which in turn determines the linear response as measured by $s_d$ to a time-dependent perturbation proportional to $s_0$. If this indeed falls off faster than $1/2^d$, then we can have some hope that the Bethe lattice can be used as a starting point for investigation of fixed valence random lattices.


\section*{Acknowledgements}
We would like to thank Sam Gutmann, Guifre Vidal, Bruno Nachtergaele, Frank Verstraete, Senthil Todadri, Subir Sachdev and Mehran Kardar for stimulating discussions. 
DN, EF and JG gratefully acknowledge support from the W. M. Keck Foundation Center for Extreme Quantum Information Theory, and from the National Security Agency (NSA) and the Disruptive Technology Office (DTO) under Army Research Office (ARO) contract W911NF-04-1-0216.
PS gratefully acknowledges support from the W. M. Keck Foundation Center for Extreme Quantum Information Theory, and from the National Science Foundation through grant CCF0431787.


\end{document}